\documentclass[10pt, conference, letterpaper]{IEEEtran}
\usepackage{comment}
\usepackage[latin1]{inputenc}
\usepackage[T1]{fontenc}

\def\BibTeX{{\rm B\kern-.05em{\sc i\kern-.025em b}\kern-.08em
    T\kern-.1667em\lower.7ex\hbox{E}\kern-.125emX}}

\usepackage{graphicx}
\usepackage{subfigure}
\usepackage{algorithm}
\usepackage{algorithmicx}
\usepackage{algpseudocode}
\usepackage{amsmath}
\usepackage{url}
\usepackage{array}
\usepackage{multirow}

\begin{document}
\title{HashFlow  For Better Flow Record Collection}
\author{Zongyi Zhao, Xingang Shi, Xia Yin, Zhiliang Wang\\ zhaozong16@mails.tsinghua.edu.cn shixg@cernet.edu.cn yxia@tsinghua.edu.cn wzl@cernet.edu.cn\\ 
	Tsinghua University}
\maketitle
\begin{abstract}
Collecting flow records is a common practice of network operators and researchers 
for monitoring, diagnosing and understanding a network. Traditional tools like 
NetFlow face great challenges when both the speed and the complexity of the network traffic increase. 
To keep pace up, we propose HashFlow, a tool for more efficient and accurate 
collection and analysis of flow records. The central idea of HashFlow is to  
maintain accurate records for elephant flows, but summarized records for mice flows, 
by applying a novel collision resolution and record promotion strategy to hash tables. 
The performance bound can be analyzed with a probabilistic model, 
and with this strategy, HashFlow achieves a better utilization of space, and also 
more accurate flow records, without bringing extra complexity. 
We have implemented HashFlow, as well as several latest flow measurement algorithms such as 
FlowRadar, HashPipe and ElasticSketch, in a P4 software switch. 
Then we use traces from different operational networks to evaluate them.
In these experiments, for various types of traffic analysis applications, 
HashFlow consistently demonstrates a clearly better performance against 
its state-of-the-art competitors. 
For example, using a small memory of 1 MB, HashFlow can accurately record 
around 55K flows, which is often 12.5\% higher than the others.
For estimating the sizes of 50K flows, HashFlow achieves a relative error of around 11.6\%, 
while the estimation error of the best competitor is 42.9\% higher. 
It detects 96.1\% of the heavy hitters out of 250K flows with a size estimation error of 5.6\%, which is 11.3\% and 73.7\% better than  
the best competitor respectively.
At last, we show these merits of HashFlow come with almost no degradation of throughput.
\end{abstract}

\section{Introduction}
\label{section:introduction}
A proper view of the statistics and the dynamics of a network is of great importance to network management.
It enables network operators to detect and correct configuration errors, 
allocate resources and perform  traffic engineering, or detect network attacks. 
NetFlow\cite{claise_cisco_2004} is a widely used tool in network measurement and analysis.
It records traffic statistics in the form of flow records, where each record contains important 
information about a flow, for example, its source and destination IP addresses, 
start and end timestamps,  type of services, application ports, input and output ports, 
as well as the volume of packets or bytes, etc. 

A challenge in implementing NetFlow like tools is to keep up with the ultra high speed 
of network traffic, especially on high-bandwidth backbone links. For example, assuming an 
average packet size of 700 bytes, and a 40 Gbps link, the time budget for processing 
one packet is only around 50 nano-seconds. In an extreme case where packets of 40 bytes 
arrive at a speed of 100 Gbps, the time budget will be only a few nano-seconds. 
NetFlow also faces the high diversity of the traffic, where hundreds of thousands, 
even millions of concurrent flows appear in measurement epoch. This pose stringent 
pressure on the scarce high speed memories, such as on-chip SDRAM with 1 $\sim$ 10 nano-seconds 
access delay\cite{li_flowradar:_2016}\cite{noauthor_access_nodate}.

One straightforward solution is to use sampling \cite{noauthor_sampled_nodate}, 
where out of several packets, only one of them gets processed and used to update the flow records.
However, sampling reduces processing overhead at the cost of less packets or flows being recorded, 
thus less accurate statistics that can be estimated. To remedy this, very enhanced sampling algorithms
\cite{hohn_inverting_2003}\cite{duffield_estimating_2005}\cite{tune_towards_2008}
have been proposed and tailored for specific measurement requirement, 
and their impact analyzed\cite{duffield2004}\cite{SamplingImpact}.
Another direction of solution is to use sketch (also referred as data streaming algorithms)
\cite{DataStreams2005}\cite{huang_sketchvisor:_2017}\cite{chen_counter_2017},
where a succinct data structure is designed and can be updated very efficiently. 
However, these sophisticated data structures and algorithms generally can only be used in limited scenarios, 
but not for the wide range of applications that the original NetFlow can support.

Towards accelerating flow record maintenance and achieving better statistics estimation, 
recently a few algorithms that make enhancement to a naive hash table and integrate sketches 
have been proposed, including OpenSketch\cite{yu_software_2013}, 
UnivMon\cite{liu_one_2016}, FlowRadar\cite{li_flowradar:_2016}, HashPipe\cite{sivaraman_heavy-hitter_2017},  
and ElasticSketch\cite{yang_elastic_2018}, etc. Both constant bound of worst case delay and 
efficient utilization of memory are achieved, making them good candidates for general 
measurement applications in high speed environment.

Following these efforts, we propose HashFlow, which makes a further step in squeezing memory consumption. 
The central idea of HashFlow is to  
maintain accurate records for elephant flows (i.e., flows with many packets), as well as summarized records for mice flows (i,e., flows with few packets), 
by applying a novel collision resolution and record promotion strategy to hash tables. 
The collision resolution part eliminates collisions that may mix up packets 
from different flows, keeps a flow from being evicted before all its packets 
have arrived, and fills up nearly all hash table buckets.
On the other hand, the promotion part bounces a flow back from the summarized set to the accurate set, when 
this flow becomes an elephant, and replaces the original one which has smaller size. 
The performance bound can be analyzed with a probabilistic model, 
and with this strategy, HashFlow achieves a better utilization of space, 
and also more accurate flow records, without bringing extra complexity. 

We have implemented HashFlow, as well as several latest flow measurement algorithms mentioned above, 
including FlowRadar, HashPipe and ElasticSketch, in a P4-programmable \cite{bosshart_p4:_2014} software switch\cite{noauthor_bmv2:_2018}. 
We then use traces from different operational networks to evaluate their effectiveness.
In these experiments, for various types of traffic analysis applications, 
HashFlow demonstrates a consistently better performance against 
its state-of-the-art competitors. 
For example, using a small memory of 1 MB, HashFlow can accurately record 
around 55K flows, which is often 12.5\% higher than the others.
For estimating the sizes of 50K flows, HashFlow achieves a relative error of around 11.6\%, 
while the estimation error of the best competitor is 42.9\% higher. 
It detects 96.1\% of the heavy hitters out of 250K flows with a size estimation error of 5.6\%, which is 11.3\% and 73.7\% better than  
the best competitor respectively.
At last, we show that these merits of HashFlow come with almost no degradation of throughput.

The remainder of the paper is organized as follows.
We introduce our motivation and central ideas in designing HashFlow in Section~\ref{section:background}. 
Then we present the algorithm details, as well as the theoretical analysis in Section~\ref{section:algorithmoverview}. 
Using real traffic traces, we analyze the parameters of HashFlow and compare it against other algorithms in Section~\ref{section:evaluation}.
Finally we conclude the paper in Section~\ref{section:conclusion}.

\section{Background and Basic Ideas}
\label{section:background}
Formally, we define a flow record as a key-value pair $(key, count)$, 
where $key$ is the ID of the flow, and $count$ is the number of packets belonging to this flow. 
A simple example is like this: the flow ID contains the source and destination IP addresses, 
while packets with exactly the same source and the same destination belong to the same flow. 
The definition is general, since the flow ID can also be a subnet prefix, a transport layer port, 
or even a keyword embedded in application data. A naive method to maintain flow records is to 
save them in a hash table, but multiple flows may be hashed to the same bucket in the table. 
Mechanisms to resolve collisions in hash tables include classic ones like separate chaining 
and linear probing, and more sophisticated ones like Cuckoo hashing \cite{pagh_cuckoo_2004}.
However, in the worst case, they need unbounded time for insertion or lookup, thus are not 
adequate for our purpose. 

Before presenting HashFlow, we briefly introduce the mechanisms of several recently proposed algorithms,
i.e., HashPipe\cite{sivaraman_heavy-hitter_2017}, ElasticSketch\cite{yang_elastic_2018} and FlowRadar\cite{li_flowradar:_2016}.
In doing this, we try to point out some minor defects in their algorithms, not for criticism, but for possible 
enhancement and new strategy we may introduce to HashFlow. 

HashPipe\cite{sivaraman_heavy-hitter_2017} uses a series of independent hash tables (each with a different hash function).
When a packet comes to the first table, if in the bucket it is hashed into, there already exists a record of another flow, 
then the old flow will be evicted to make room for the new flow. The evicted flow record will then try to find an empty bucket 
in the remaining hash tables as a newcomer. When a collision happens there, among the newcomer and the existing record, 
the one with a smaller packet count will be kicked out, and become the newcomer to the next table. This process goes on 
until either an empty bucket is found, or there is no remaining hash table (in which case the last evicted flow will be discarded). 

HashPipe uses the first table to effectively accommodate new flows and evict the existing flows when collision occurs, otherwise new flows will have little chance to stay if large flows accumulate in the table. But on the other hand, this strategy frequently 
splits one flow record into multiple records that are stored in different hash tables, each with a partial count, 
since an existing flow may be evicted but new packets of this flow may still arrive later. 
This effect makes the utilization of memory less efficient, and makes the packet count less accurate. 

ElasticSketch\cite{yang_elastic_2018} uses, in the hash table, 
two packet counters (vote$^{+}$ and vote$^{-}$) instead of one in each flow record. 
Vote$^{+}$ maintains the number of packets belonging to the flow, while vote$^{-}$ for the packets belonging to all the 
other flows that have been hashed into the same bucket. It also uses a count-min sketch\cite{cormode_countmin_2005} 
to maintain summarized flow records, corresponding to both flows that have never been stored in the hash table, 
and flows that have been evicted when the corresponding vote$^{-}$ is too large, i.e., $\frac{vote^{-}}{vote^{+}}$ 
is greater than a predefined threshold $\lambda$. 
However, due to collisions and the eviction strategy ElasticSketch employs, a flow record may also be split 
into multiple records, and the packet counter is not accurate. The count-min sketch is introduced to help the flow size 
estimation. However, since the count-min sketch itself may not be accurate, the estimation accuracy is limited,  
which is especially true if the sketch is occupied by too many flows.

FlowRadar\cite{li_flowradar:_2016} uses a bloom filter\cite{bloom_space/time_1970} to determine 
whether a new flow comes in. It also adds a flow count and a flow set field to each flow record in the hash table. 
For each packet, FlowRadar hashes it into multiple buckets, and updates the corresponding packet count fields. 
If a new flow comes in as reported by the bloom filter, the flow set fields of the buckets will be ``xor''-ed with the new ID, 
and the flow count fields will be incremented by 1. 
During post processing phase, FlowRadar can decode (some) flow IDs that are encoded into the flow set fields, and also recover the corresponding packet counts. 
However, the chances that such decoding succeeds drop abruptly if the table is heavily loaded and there are not enough flows that don't collide with any other ones.



With these in mind, we then analyze a few tradeoffs and design choices 
for time and space efficient collection of flow records.

1) {\em With Limited memory, discard flows when necessary.} Pouring too many flows into a 
hash table or sketch will cause frequent collision, and either increase the processing overhead, 
or decrease the accuracy of the information that can be retrieved. 
For example, FlowRadar faces severe degradation in its decoding capability
 when the number of flows exceeds its capacity 
(this effect and the turning point can be clearly seen in our evaluation, 
for example, Fig. \ref{fig:comparison_concurrent_flows_increases_flow_monitoring} for flow set monitoring 
and Fig. \ref{fig:comparison_concurrent_flows_increases_fs_estimation} for flow size estimation).  
In most situations, network traffic is skewed such that 
only a small portion of elephant flows contain a large number of packets. 
For example, in one campus trace we use, 7.7\% of the flows contribute more than 85\% of the packets. 
It will be better to discard mice flows with few packets than elephant ones, 
since the latter have a greater impact on most applications, 
such as heavy hitter detection, traffic engineering and billing. It is often enough to maintain summarized information for the mice flows.

2) {\em A flow should be consistently stored in one record.} Both HashPipe and ElasticSketch 
may split one flow into multiple fragments stored in different tables. This not only wastes memory, 
but also causes the packet count less accurate, which in turn affects the eviction strategies.  
By storing a flow in a consistent record, we can achieve both better memory utilization and 
higher accuracy. 

3) {\em If better memory utilization can be achieved by trading off a little efficiency, have a try.} 
This is particularly worth to do when network equipment is becoming more ``soft defined'', 
where their functionalities can be  ``programmed'', 
and the additional operations can be easily paralleled or pipelined. 
By the nature of the ball and urn model \cite{urn} of hash tables, 
there will be a few empty buckets of a hash table that have never been used, and the utilization will be improved by feeding more flows into the hash table or hashing a flow multiple times to find a proper bucket. 
Both HashPipe and ElasticSketch propose to split the hash table into multiple small tables and use multiple hash functions to improve the utilization, 
but in different ways. Our collision resolution strategy is more similar to 
that of HashPipe than ElasticSketch. 
Later, we will show our strategy can make an effective use of the table buckets.

\section{Algorithm Details}
\label{section:algorithmoverview}
In this section we explain how HashFlow works in detail, and present some theoretical analysis results, 
which are based on a probabilistic model.

\subsection{Data Structures and Algorithm}
The data structure of HashFlow is composed of a main table (${\mathbf M}$) and an ancillary table (${\mathbf A}$), 
each being an array of buckets, and each bucket (also called as cell) can store a flow record
in the form of \emph{(key, count)}, as mentioned in Sec \ref{section:background}. 
In the main table ${\mathbf M}$, flow ID will be used as \emph{key}, 
while in the ancillary table ${\mathbf A}$, a digest of flow ID will be used as \emph{key}.
We have a set of $d+1$ independent hash functions, i.e., $h_1, h_2, \cdots, h_d$, and $g_{1}$, 
where $d$ is a positive integer (we  call $d$ the  \emph{depth} of ${\mathbf M}$, and typically $d=3$). 
Each hash function $h_{i} (1 \le i \le d)$  randomly maps a flow ID to one bucket in ${\mathbf M}$, 
while $g_{1}$ maps the ID to one bucket in ${\mathbf A}$. 
A digest can be generated from the hashing result of the flow ID with any $h_i$.
When a packet arrives, HashFlow updates ${\mathbf M}$ and ${\mathbf A}$ with the following 
two strategies, as shown in Algorithm \ref{alg: process_packet}.

1) \textbf{Collision Resolution.} When a packet $p$ arrives, we first map it into the bucket indexed 
at $idx=h_1(p.\text{flow\_id})$ in the main table ${\mathbf M}$.  
If ${\mathbf M}[idx]$ is empty, we just put the flow ID and a count of 1 in the bucket (line 5$\sim$ 6).
If the bucket is already occupied by packets of this flow earlier, 
we just simply increment the count by 1 (line 7 $\sim$ 8). 
In either case, we have found a proper bucket for the packet, and the process finishes. 
If neither of the two cases happen, then a collision occurs, 
and we repeat the same process but with $h_2, h_3, \cdots, h_d$ one by one, 
until a proper bucket is found for the packet. 
This is a simple collision resolution procedure. 
Unlike HashPipe and ElasticSketch, it does not evict existing flow record from the main table, 
thus prevents a record from being split into multiple records. 

If collision cannot be resolved in the main table, 
then we try to record it in the ancillary table ${\mathbf A}$.
Here the action is more intrusive, 
as an existing flow will be replaced (discarded) if it collides with the new arrival (line 16 $\sim$ 19).

2) \textbf{Record promotion.} If, in the ancillary table ${\mathbf A}$, $p$ succeeds to find the right bucket to reside, 
then it updates the packet count field of the record. 
However, if the corresponding flow record keeps growing and becomes elephant, 
i.e., with the packet count becomes large, 
then we will promote the record by re-inserting it to the main table,
thus prevents large flows from being discarded. 
To implement this strategy, we keep in mind the sentinel flow record that has the smallest packet count 
among those records that collide with $p$ in the collision resolution procedure (line 9 $\sim$ 11). 
When a flow record in the ancillary table should be promoted, it will replace this sentinel  
we have kept in mind (line 22 $\sim$ 23).
We note that, instead of the flow ID, a shorter digest is used as keys in the ancillary table to 
reduce memory consumption. This may mix flows up, but with a small chance.

\begin{algorithm}[ht!]
    \caption{Update Algorithm of HashFlow on arrival of $p$}
    \label{alg: process_packet}
    \algrenewcommand\algorithmicwhile{\textbf{when}}
    \begin{algorithmic}[1]
        \State{//Collision Resolution}
        \State{$flowID \gets p.\text{flow\_id}, min \gets \infty, pos \gets -1$}
        \For{$i=1$ to $d$}
        \State{$idx\gets h_{i}(flowID)$}
        \If{${\mathbf M}[idx].count==0$}
        \State{${\mathbf M}[idx] \gets (flowID, 1)$}
        \Return
        \ElsIf {${\mathbf M}[idx].key == flowID$}
        \State{Increment ${\mathbf M}[idx].count$ by 1}
        \Return
        \ElsIf{${\mathbf M}[idx].count < min$}
        \State{$min \gets {\mathbf M}[idx].count$}
        \State{$pos \gets idx$}
        \EndIf
        \EndFor
        \State{$idx\gets g_{1}(flowID)$}
        \State{$digest \gets h_{1}(flowID)\%(2^{\text{digest width}})$}
        \If{${\mathbf A}[idx].count==0$ or ${\mathbf A}[idx].key \neq digest$}
        \State{${\mathbf A}[idx] \gets (digest, 1)$}
        \ElsIf{${\mathbf A}[idx].count < min$}
        \State{Increment ${\mathbf A}[idx].count$ by 1}
        \Else
        \State{//Record Promotion}
        \State{${\mathbf M}[pos].key \gets flowID$}
        \State{${\mathbf M}[pos].count \gets {\mathbf A}[idx].count+1$}
        \EndIf
    \end{algorithmic}
\end{algorithm}

\begin{figure}
    \centering
    \includegraphics[width=0.9\linewidth]{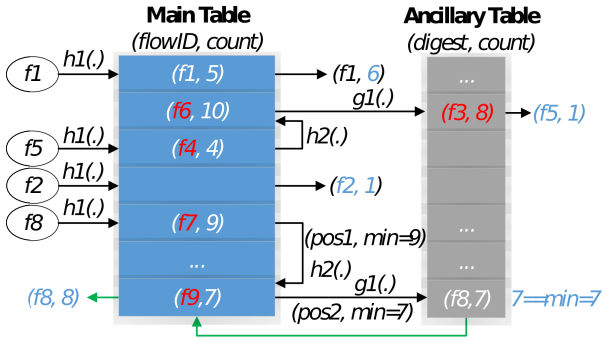}
    \caption{An example of HashFlow}
    \label{fig:datastructure}
\end{figure}

We use a simple example with $d=2$ to illustrate the algorithm, as depicted in Fig. \ref{fig:datastructure}.
When a packet of flow $f_1$ arrives, $h_1$ maps it into a bucket indexed at $h_1(f_1)$, 
where the record$(f_{1}, 5)$ has the same key, 
and the counter is simply incremented. 
When a packet of flow $f_2$ arrives, $h_1$ maps it into an empty bucket, 
so the record becomes $(f_{2}, 1)$. 
When a  packet of flow $f_5$ arrives, it collides with the record $(f_4, 4)$ in the bucket indexed at $h_1(f_5)$. 
Then we try to resolve collision with $h_2$, but again, 
the packet collides with the record $(f_6, 10)$ in the bucket at $h_2(f_5)$. 
So we have to use $g_1$ to find a place in the ancillary table for $p$. 
Sadly, it collides again with the record $(f_3, 8)$, and we let it replace the existing one.
The last packet is from  flow $f_{8}$, and it goes through a similar process to that of $f_5$. 
The difference is that, at last, this packet finds its corresponding flow record of $(f_8,7)$ in the ancillary table, and the record becomes elephant 
(as the sentinel flow with the smallest packet count in the main table has a packet count of 7). 
So we promote $(f_8, 8)$ by inserting it back into the main table, evicting the sentinel one.

In Algorithm \ref{alg: process_packet}, we use multiple independent hash functions in the main table ${\mathbf M}$.
Another choice is to use multiple small hash tables ${\mathbf M}_{i}$, each of which corresponds to the  hash function $h_{i}$. 
Algorithm \ref{alg: process_packet} can be modified straightforwardly: 
to update the $i$-th small table ${\mathbf M}_{i}$ instead of ${\mathbf M}$ (line 5 $\sim$ 12),
to remember which small table the sentinel record resides in (line 10 $\sim$ 11), 
and to evict the sentinel record in the right small table (line 22 $\sim$ 23).
In addition, we introduce a weight $\alpha (0<\alpha<1)$, 
such that the number of buckets in $\mathbf{M}_{i+1}$ is $\alpha$ times that of $\mathbf{M}_i$.

Readers may notice that HashPipe uses a similar scheme of pipelined tables, 
but there are a few important differences. 
First, HashFlow uses pipelined tables together with an ancillary table.
Second, the update strategy of these pipelined tables is different from that of HashPipe. 
Third,  our collision resolution procedure on the main table can be analyzed theoretically, 
based on which we can achieve a concrete performance guarantee on the number of accurate 
flow records that HashFlow can maintain.

\subsection{Analysis}
\label{analysis}
In the section, we propose a probabilistic framework that models the utilization of the  main table ${\mathbf M}$.
We first analyze the case where a multi-hash table is used for ${\mathbf M}$,
then the case where pipelined tables are used. 
In either case, we assume that there are $m$ distinct flows fed into ${\mathbf M}$, 
which has $n$ buckets in total, and uses $d$ hash functions.


\textbf{Multi-hash table.} First, consider the case when $d=1$, 
where the analysis follows a classic ball and urn problem\cite{urn}.
After inserting $m_1=m$ flows randomly into $n$ buckets, 
the probability that a given bucket is empty is 
\[\label{equation1}
p_1 = (1 - \frac{1}{n})^{m_1} \approx e^{-\frac{m_1}{n}},
\]
and the utilization of the table is 
$u_{1}=1-p_{1}= 1 - e^{-\frac{m_1}{n}}.$
Since each bucket can contain only one flow record due to our collision resolution strategy, 
the number of flows that fail to be cached in $\mathbf{M}$ after this round is $m_1-n\times(1-p_1)$.

Now consider the case of $d=2$. 
Essentially, a flow tries another bucket with $h_2$ if it finds out 
that the first bucket it tries has already been occupied. 
Since we don't care which exact flow is stored in the table, 
we slightly change the update process to the following one.
We take two rounds. In the first round, 
we feed all the $m_1$ flows into the table with $h_1$, exactly the same as $d=1$.
In the second round, we feed all the remaining flows that have not been kept in $\mathbf{M}$ into the table again, 
but this time with $h_2$. 
Assume $\mathbf{M}$ is empty before the second round starts, 
then after the $m_2=m_1-n\times(1-p_1)$ flows left by the first round have been inserted in the second round, 
a bucket will be empty with probability $e^{-\frac{m_2}{n}}$. 
However, $\mathbf{M}$ is actually not empty before the second round, 
and at that time a bucket in it is empty with probability $p_1$.
Since $h_1$ and $h_2$ are independent, we know after the second round, 
the probability that a bucket is still empty becomes $p_2 \approx p_1 \times e^{-\frac{m_2}{n}}$, 
and the number of flows that have not been inserted into $\mathbf{M}$ will be $m_3=m_1-n\times(1-p_2)$.
The utilization of $\mathbf{M}$ now becomes $u_2=1-p_2$.

The analysis for the slightly changed process can be extended to cases when $d>2$. 
In the $k$-th round, $m_k$ flows are fed into a hash table with a new hash function $h_k$, 
where there are already $n \times (1-p_{k-1})$ buckets being occupied in the previous rounds. 
Then after the $k$-th round, the probability that a bucket is empty is
\begin{eqnarray}
\label{xx1}
p_k & \approx &p_{k-1} \times e^{-\frac{m_k}{n}} \nonumber \\
& = & p_{k-1} \times e^{-\frac{m_1-n \times (1-p_{k-1})}{n}} \nonumber \\
& = & p_{k-1} \times e^{1-\frac{m_1}{n}-p_{k-1}} \nonumber \\
& = & p_{k-1} \times e^{1-\frac{m}{n}-p_{k-1}}
\end{eqnarray}
for $k \ge 2$. With Equation (\ref{xx1}), for any given $d$, $m$, and $n$, 
we can recursively compute the probability $p_d$ that a bucket is empty in the hash table after $d$ rounds. 
Then the utilization of the hash table will be $1-p_d$. 
We note that there is a slight difference between this model and our multi-hash table, as will be shown later.

\begin{figure*}[t]
    \centering
    \mbox{
        \subfigure[Multi-hash Table\label{multihash}]{\includegraphics[width=0.24\linewidth]{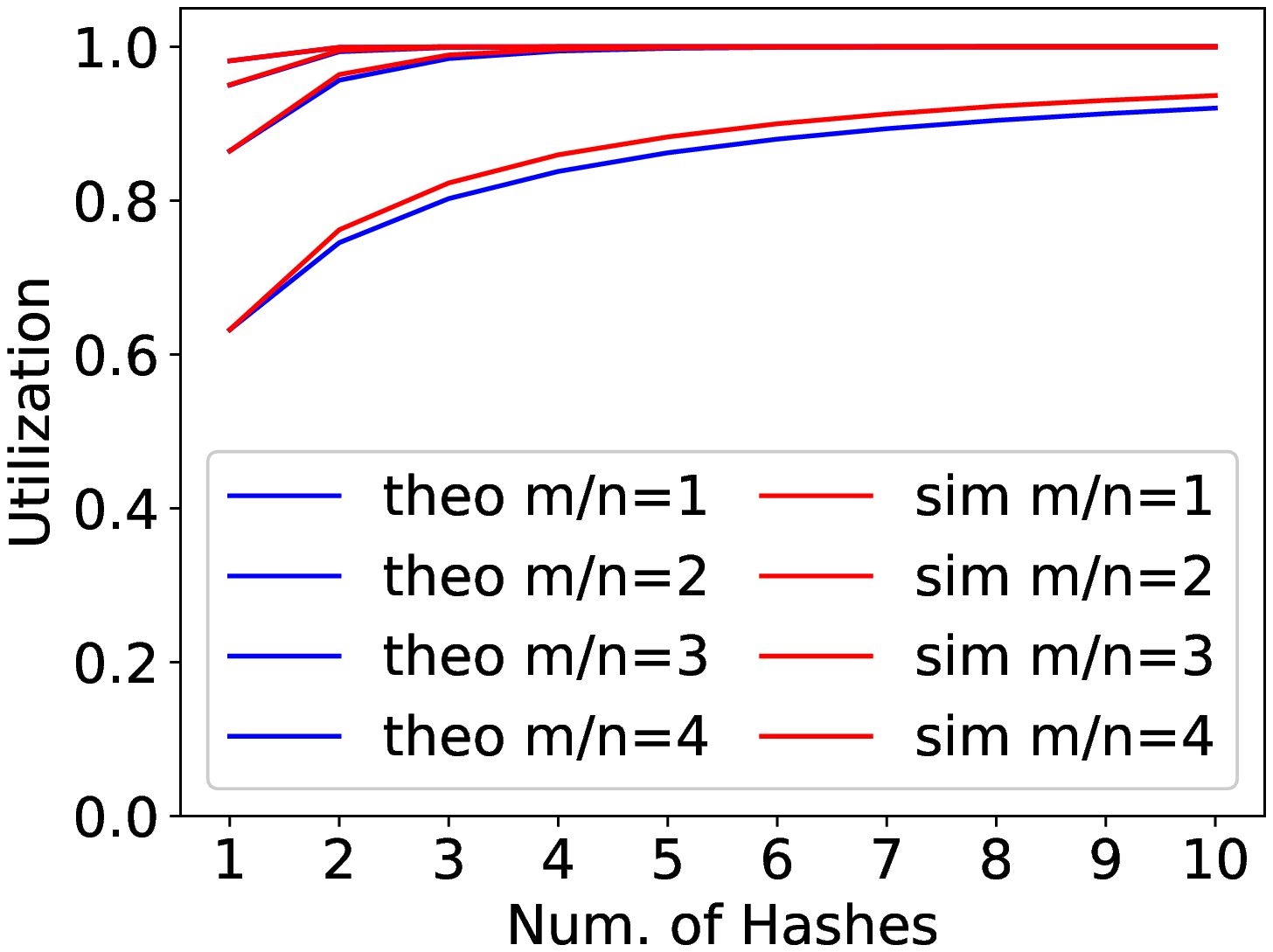}}
        \subfigure[Pipelined Tables\label{pipeline1}]{\includegraphics[width=0.24\linewidth]{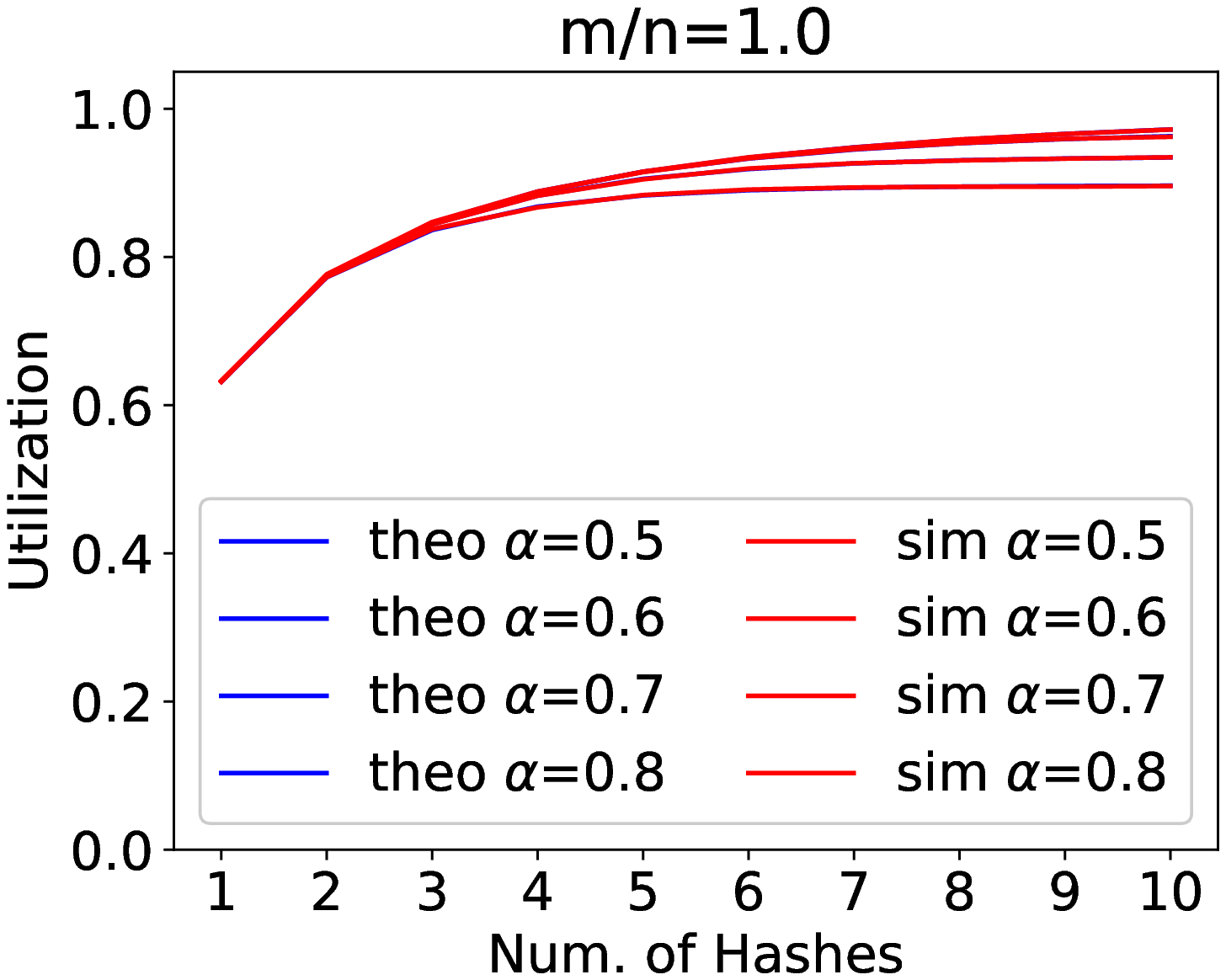}}
        \subfigure[Pipelined Tables\label{pipeline2}]{\includegraphics[width=0.24\linewidth]{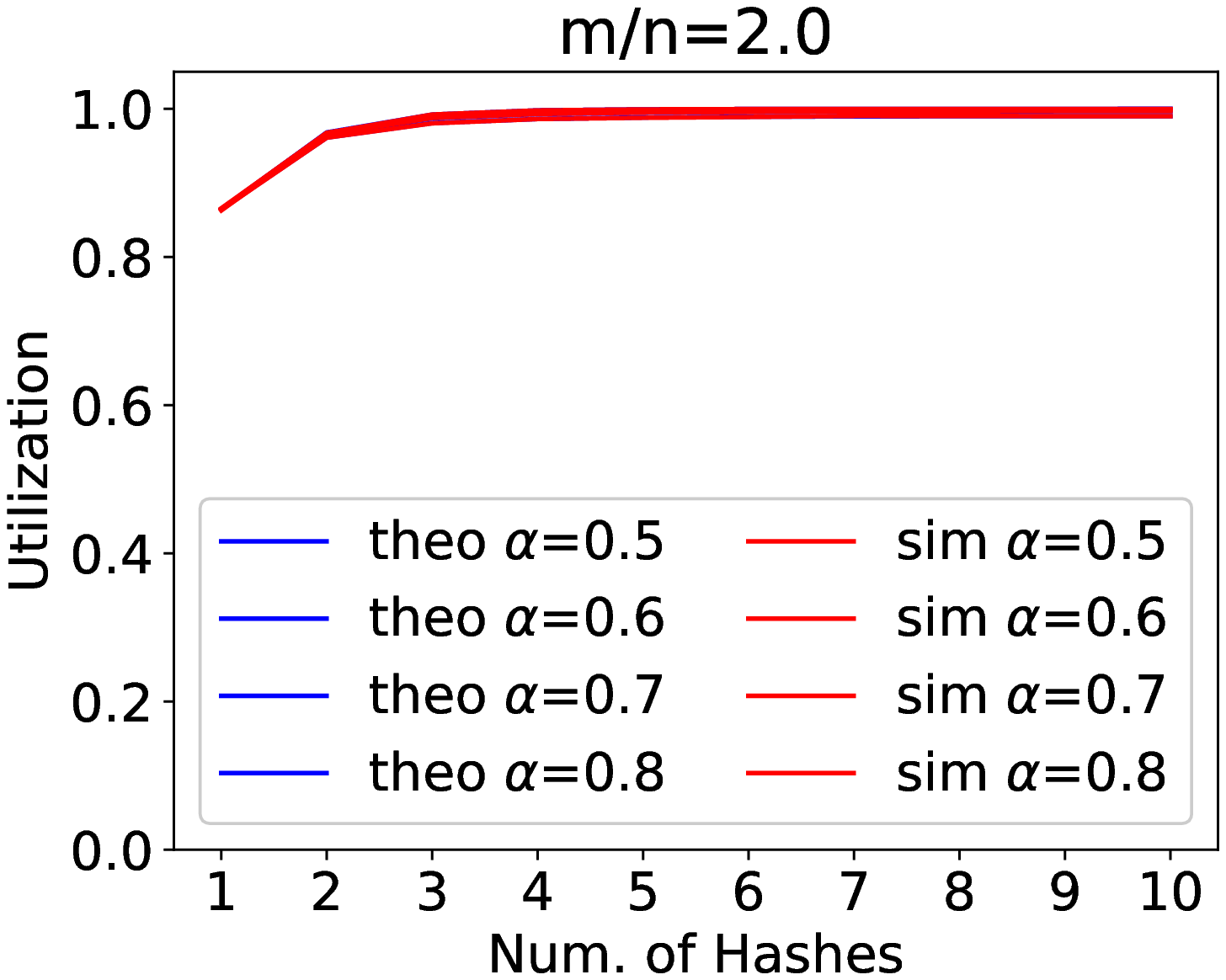}}
        \subfigure[Improvement on Utilization\label{improvement}]{\includegraphics[width=0.24\linewidth]{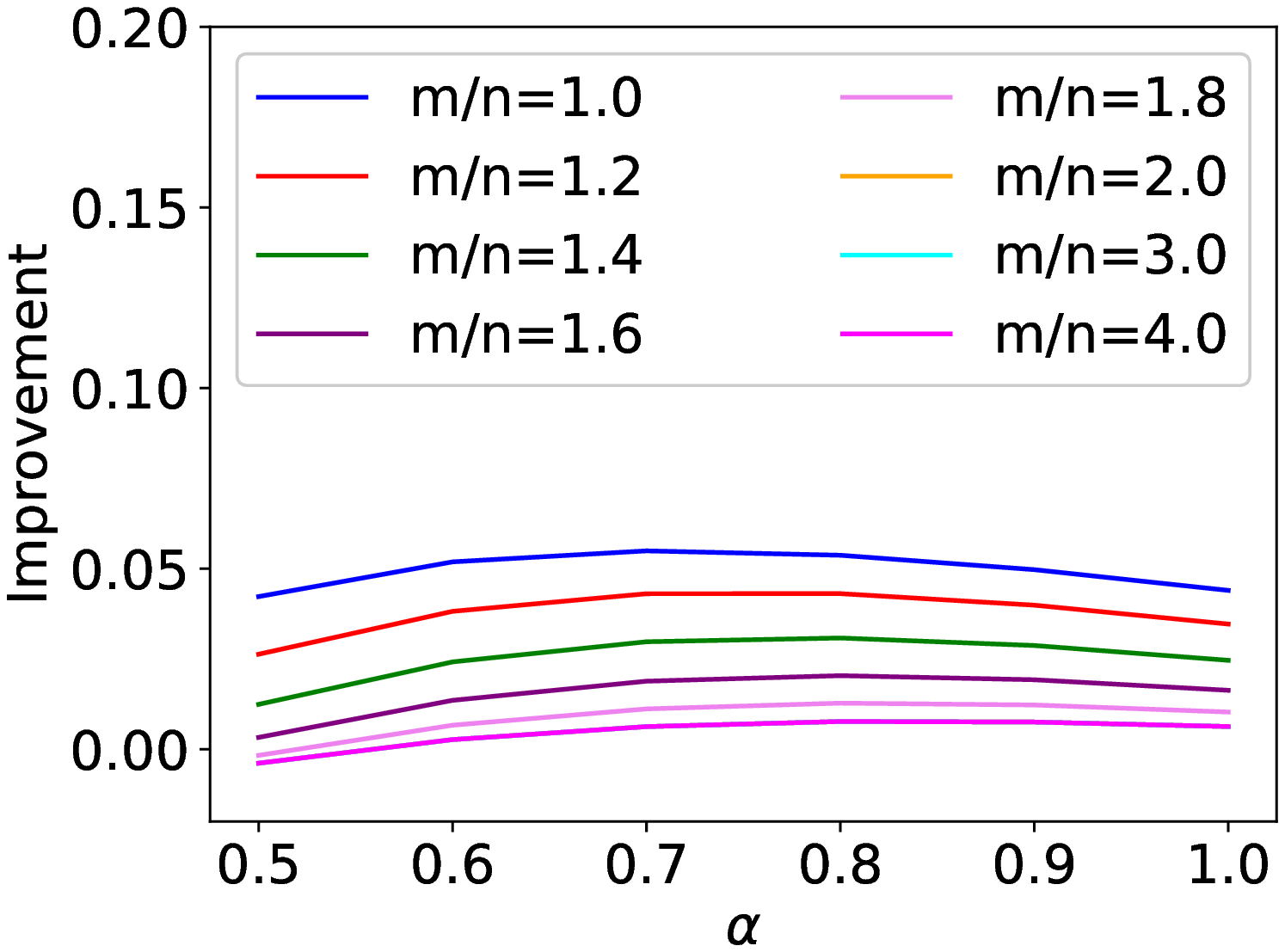}}
    }
    \caption{Utilization of the multi-hash table and the pipelined tables}
    \label{fig:pipelinedtablesutilizationratio}
\end{figure*}

\textbf{Pipelined tables.}
Let $n_k$ be the number of buckets in the $k$-th table $\mathbf{M}_k$ such that $n_{k+1}=\alpha \times n_k$, 
where $\alpha$ is the  pipeline weight.
We perform a similar modification to our collision resolution procedure with pipelined tables, 
such that in the $k$-th round, 
all packets goes though the $k$-th table before they are fed into the $k+1$-th table in the $k+1$-th round.
We use the same notations $p_k$, $m_k$ and $u_k$ as those in the first model.
Since $\sum_{k=1}^dn_k=\sum_{k=1}^d(\alpha^{k-1} \times n_1)=\frac{1-\alpha^{d}}{1-\alpha} \times n_1$, 
we get $n_1=\frac{1-\alpha}{1-\alpha^{d}} \times n$, 
and $n_k=\alpha^{k-1} \times \frac{1-\alpha}{1-\alpha^{d}} \times n$.

The first round works exactly the same as that in the previous model, 
so we get $p_{1}=(1 - \frac{1}{n_1})^{m_1} \approx e^{-\frac{m_1}{n_{1}}}$, 
$u_1=1-p_1$, and $m_2=m_1-n_{1}\times(1-p_1)$.

For the $k$-th round, we know $m_k$ flows are to be fed into the table $\mathbf{M}_{k}$ with $n_k$ buckets, 
so we get $p_{k} \approx e^{-\frac{m_k}{n_{k}}}$, 
and the number of flows left after this round is 
\begin{equation}
\label{xxx1}
m_{k+1}  =  m_{k}-n_{k}\times(1-p_k). 
\end{equation}
Dividing  both sides of Equation (\ref{xxx1}) by $n_{k+1}$, we get 
\begin{eqnarray}
\frac{m_{k+1}}{n_{k+1}} & = & \frac{n_k}{n_{k+1}}  \times \frac{m_{k}-n_{k}\times(1-p_k)}{n_k} \nonumber \\
& = & \alpha^{-1} \times \left( \frac{m_k}{n_k} - 1 + p_k  \right), \nonumber
\end{eqnarray}
which is just
\begin{equation}
\label{xxx2}
-\ln p_{k+1}=\alpha^{-1} \times (-\ln p_k -1 + p_k).
\end{equation}

From Equation (\ref{xxx2}), we finally get 
\begin{equation}
\label{xxx3}
p_{k+1}=(p_k)^{\frac{1}{\alpha}} \times e^{\frac{1-p_k}{\alpha}}. 
\end{equation}

With Equation (\ref{xxx3}), for any given $d$, $m$, and $n$, 
we can recursively compute the probability $p_k (1 \le k \le d)$ that a bucket is empty in the $k$-th hash table.
Then the utilization of the pipelined tables will be 
\begin{equation}
\label{pipelineutil}
\frac{\sum_{k=1}^{d} (n_k \times (1-p_k))}{\sum_{k=1}^{d} n_k} = 1- \frac{1-\alpha}{1-\alpha^{d}} \times \sum_{k=1}^{d}(\alpha^{k-1} \times p_{k}).
\end{equation}

Now we will show how accurate the models are. 
In Fig. \ref{multihash},
we compare the utilization provided by our multi-hash table model against the results from simulations on some real traces.
We use $n=$100K buckets, with different depth $d$ from 1 to 10, 
and vary the traffic load $m/n$ from 1 to 4. As can be seen there, 
only under a light load of $m/n=1$, there is a slight difference between the model and the real algorithm.
When $m/n \ge 2$, the multi-hash table model provides nearly perfect predictions.

Fig. \ref{pipeline1} and Fig. \ref{pipeline2} depict the utilization provided by our model on pipelined tables, 
as well as results from simulations, for traffic load $m/n=1.0$ and 2.0, respectively. 
We use a similar setting as above, with $n=$100K and $d$ from 1 to 10,
but we vary the pipeline weigh $\alpha$ between 0.5 to 0.8. 
This time the model and the simulation results match  quite well, 
since  arranging the packet arrivals in rounds, as we have done in the model,  
actually does not affect the final probability (we omit the proof due to space limitations).

With these two models, we can compute the utilization of our main table, 
as long as the traffic load $m/n$ is known. Since each record is accurate 
(neglecting the minor chance that a flow record promoted back to the main 
table happens to have an inaccurate count), this provides a concrete prediction 
on the number of records HashFlow can report. 
We can see  more hash functions will improve the utilization.
For example, in the case of $m/n=1$, the utilization  increases from 63\% to 80\% 
when $d$ is increased from 1 to 3, and from 83 to 92 when $d$ is increased from 3 to 10.
As more hash functions require more hash operations and memory accesses in the worst case, 3 hash functions seems to be a  sweet spot, and we use $d=3$ by default in our evaluations.

Fig. \ref{improvement} shows, when $d=3$, 
pipelined tables always improves the utilization upon  multi-hash table, regardless of the traffic load.
As shown there, when $\alpha=0.7$ and $m/n=1$, up to 5.5\% more utilization can be achieved.
Our evaluation will adopt the pipelined scheme, where $\alpha=0.7$ seems to be the best choice.


\section{Evaluation}
\label{section:evaluation}

\subsection{Methodology}\label{methodology}
We have implemented HashFlow, as well as several latest algorithms that try to improve NetFlow, 
including  FlowRadar\cite{li_flowradar:_2016}, HashPipe\cite{sivaraman_heavy-hitter_2017} 
and ElasticSketch\cite{yang_elastic_2018}, in bmv2 \cite{noauthor_bmv2:_2018}, 
which is a software switch with P4 \cite{bosshart_p4:_2014} programmability. 
The code for FlowRadar and ElasticSketch are rewritten based on their published code, 
while HashPipe is implemented based on the algorithm in the published paper. 

\begin{figure}
\centering
\begin{minipage}{.45\linewidth}
    \centering
    \includegraphics[width=\linewidth]{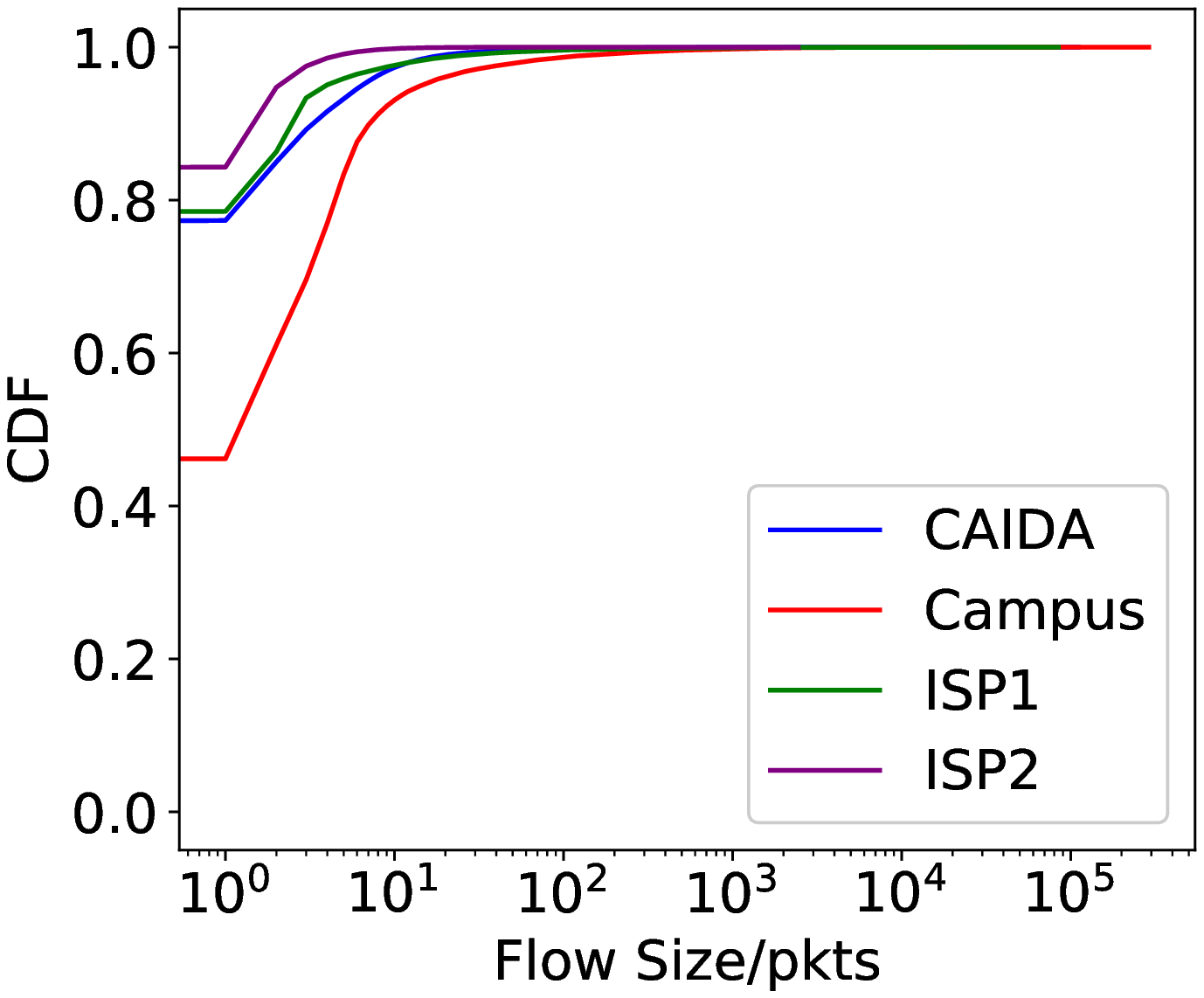}
    \caption{Flow size distribution of the traces used for evaluation}
    \label{fig:flowsizedistribution}
\end{minipage}
\begin{minipage}{.45\linewidth}
	\centering
	\includegraphics[width=\linewidth]{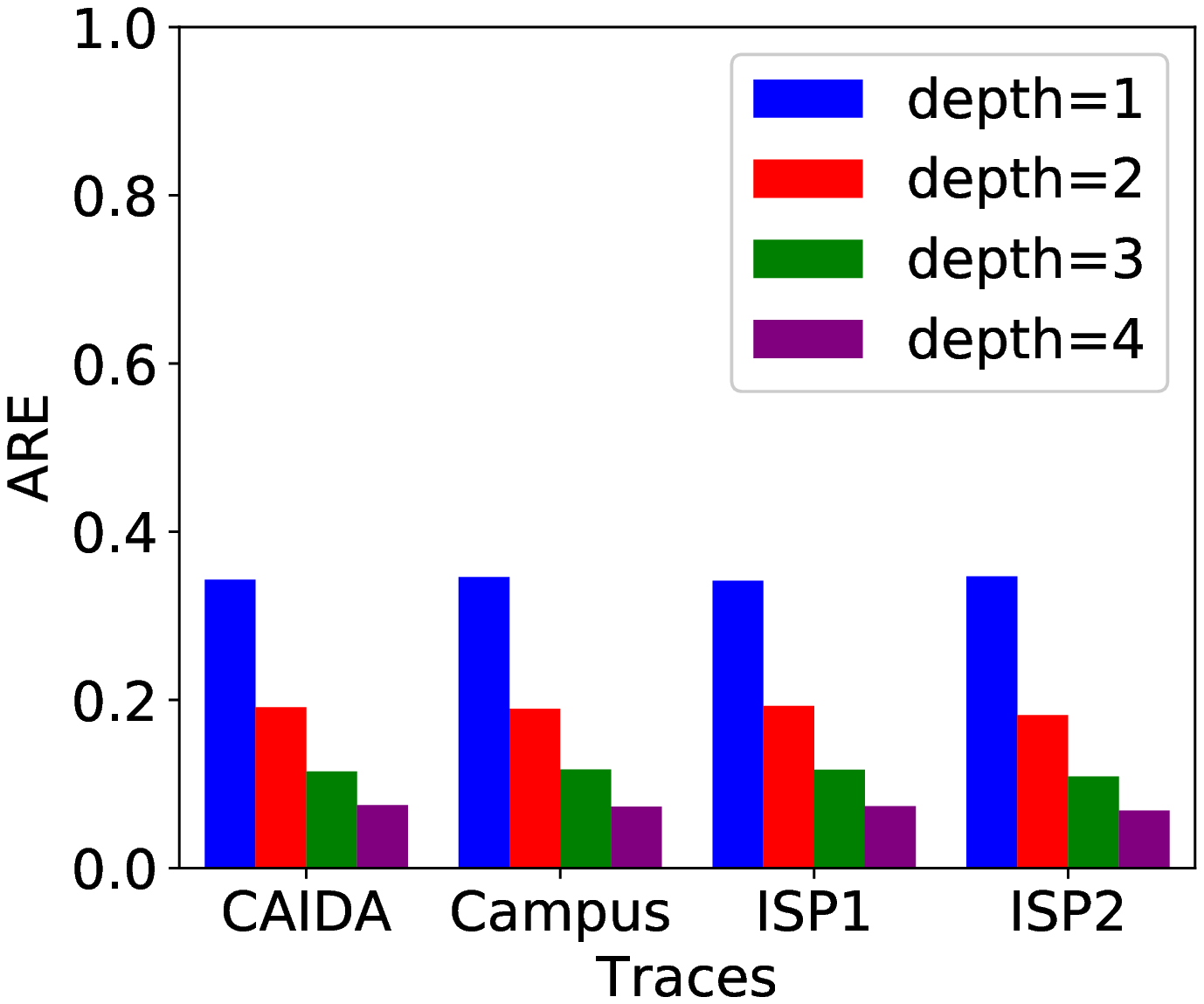}
	\caption{Flow size estimation under different pipeline depth}
	\label{fig:comparison_increase_depth}
\end{minipage}
\end{figure}

We use 4 traces from different environment to evaluate these algorithms' performance, 
one from a 40 Gbps backbone link provided by CAIDA \cite{noauthor_caida_nodate},
one from a 10 Gbps link in a campus network, and the other two from different ISP access networks.
Some flow level statistics are summarized in Table~\ref{tab:netflowtraces}, 
where we can see the traffic in different traces differ greatly.
However, by plotting the cumulative flow size distribution in Fig. \ref{fig:flowsizedistribution}, 
we can find they all exhibit a similar skewness pattern, 
that most flows are mice flows with a small number of packets, 
while most of the traffic are from a small number of elephant flows \cite{benson_network_2010}.
The only exception is the ISP2 trace, which is 1:5000 sampled from an access link and 
more than 99\% of the flows in it have less than 5 packets (the CDF also reveals this).
When evaluating the algorithms, for each trial, we select a constant number of flows from each trace, 
and feed the packets of these flows to each algorithm.

\begin{table}
    \centering
    \caption{Traces used for evaluation}
    \label{tab:netflowtraces}
    \begin{tabular} {c | c | c | c }
    \hline\hline
    Trace & Date & max flow size & ave. flow size \\
    \hline
    CAIDA &2018/03/15&110900 pkts & 3.2 pkts\\
    Campus &2014/02/07&289877 pkts & 15.1 pkts\\
    ISP1 &2009/04/10&84357 pkts& 5.2 pkts\\
    ISP2 &2015/12/31&2441 pkts& 1.3 pkts\\
    \hline
    \end{tabular}
\end{table}

Suppose $n$ flows are processed by each algorithm.
The measurement applications we use to evaluate the algorithms 
and traffic statistics we use as performance metrics are as follows. 

\begin{itemize}
    \item \emph{Flow Record Report.} 
An algorithm reports the flow records it maintains, 
where each record is of the form (flow ID, packet count). 
The performance metric we  use is \emph{Flow Set Coverage (FSC)} defined as 
\[FSC\!=\!\frac{\text{num. of flow records with correct flow IDs}}{n}.\]
Notice that HashPipe, ElasticSketch and HashFlow maintains flow records individually, 
while FlowRadar can decode flow records from a coded flow set.

    \item \emph{Flow Size Estimation.} 
Given a flow ID, an algorithm estimates the number of 
packets belonging to this flow. If no result can be reported, we use 0 as the default value. 
The performance metric we  use is \emph{Average Relative Error (ARE)} defined as 
\[ARE=\frac{1}{n}\sum\left|\frac{\text{estimated size of flow } i }{\text{real size of flow } i}-1\right|.\]
Notice that, FlowRadar can decode flow sizes from a coded flow set, 
while ElasticSketch can use an additional count-min sketch to estimate flow sizes.

    \item \emph{Heavy Hitter Detection.} 
An algorithm reports heavy hitters, which are flows with more than $T$ packets, 
and $T$ is an adjustable parameter. 
Let $c_1$ be the number of  heavy hitters reported by an algorithm, 
$c_2$ the number of real heavy hitters, 
and among the reported $c_1$ heavy hitters $c$ of them are real.
The performance metric we use is \emph{F1 Score} defined as
\[\text{F1 Score}=\frac{2\cdot PR\cdot RR}{PR + RR},\]
where $PR = \frac{c}{c_1}$  and $RR=\frac{c}{c_2}$. 
We also use \emph{ARE} of the size estimation of the heavy hitters 
as another metric.

    \item \emph{Cardinality Estimation.} 
An algorithm estimates the number of flows. 
The performance metric we  use is \emph{Relative Error (RE)} defined as 
\[RE =\left|\frac{\text{estimated number of flows}}{n}-1\right|.\]
Notice that, linear counting\cite{whang_linear-time_1990} is used by ElasticSketch
to estimate the number of flows in its count-min sketch, 
and used by HashFlow to estimate the number of flows in its ancillary table.

\end{itemize}

Following recommendations in the corresponding papers, we set the parameters of these algorithms as follows. 
\begin{itemize}
    \item HashPipe: We use 4 sub-tables of equal size.
    \item ElasticSketch: We adopt the hardware version, where 3 sub-tables
are used in its heavy part. The light part uses a count-min sketch of one array, 
and the two parts use the same number of cells.
    \item FlowRadar:  We use 4 hash functions for its bloom filter and 3 hash functions for its counting table.
The number of cells in the bloom filter is 40 times of that in the counting table. 
    \item HashFlow: We use the same number of cells in the main table and the ancillary table. 
The main table consists of three small hash tables, while the weight $\alpha$ is 0.7 unless otherwise stated. 
Each digest and counter in the ancillary table costs 8 bits.
\end{itemize}

We let these algorithms use the same amount of memory in all the experiments. 
For each flow record, we use a flow ID of 104 bits and a counter of 32 bits,
So 1 MB memory approximately corresponds to 60K flow records.
In the worst case, HashFlow, HashPipe and ElasticSketch (hardware version) will compute 4 hash results 
to access the corresponding cells, while FlowRadar needs to compute 7 hash results.


\subsection{Optimizing the Main Table}
We first demonstrate the performance of the main table with the collision resolution strategy, 
under different settings and parameters, i.e, using a multi-hash table, 
or using pipelined tables with different weights.

\begin{figure}[t]
    \centering
    \mbox{
        \subfigure[Flow Record Report\label{MT-flowrecord}]{\includegraphics[width=0.45\linewidth]{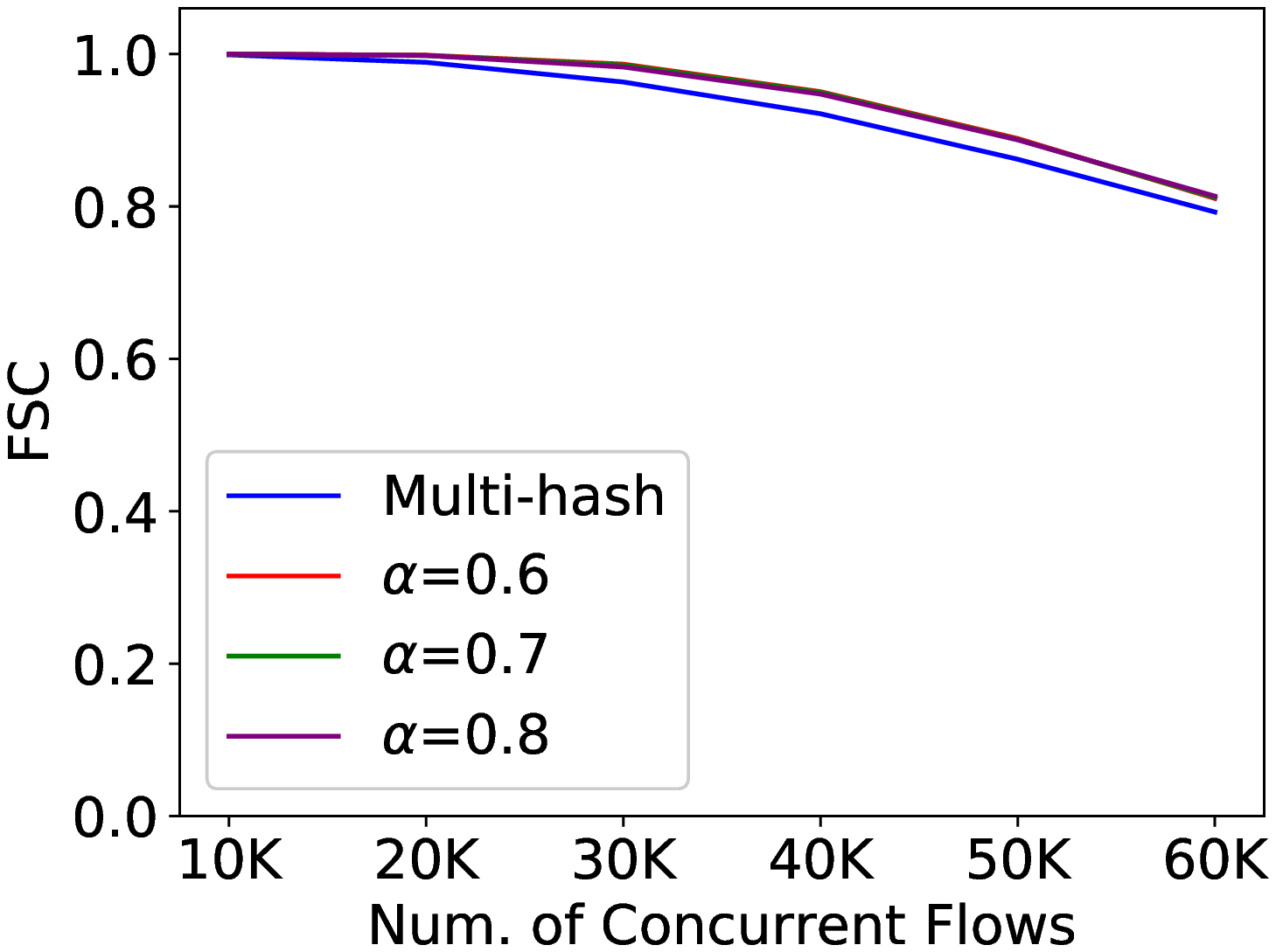}}
        \subfigure[Flow Size Estimation\label{MT-flowsize}]{\includegraphics[width=0.45\linewidth]{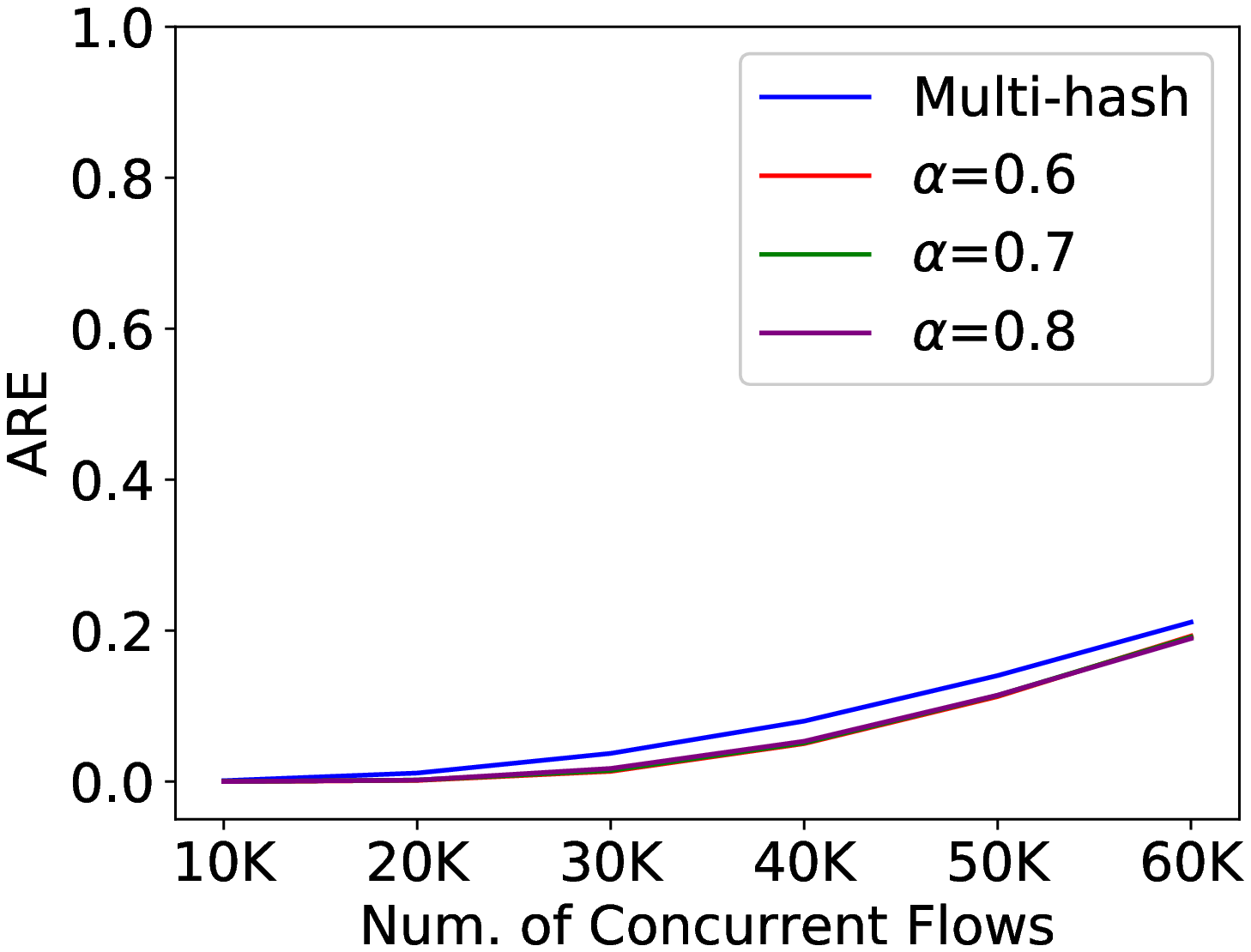}}
    }
    \caption{Comparing multi-hash table with pipelined tables.}
    \label{fig:performance_comparison_hierarchical_hashtable}
\end{figure}

We plot the \emph{Flow Set Coverage (FSC)} for flow record report in Fig. \ref{MT-flowrecord},
and plot the \emph{Average Relative Error (ARE)} for flow size estimation in Fig. \ref{MT-flowsize}, 
where there are 3 pipelined tables, and the weight is 0.6, 0.7 and 0.8 respectively. 
The traces are from  the campus network, and as the number of flows increases from 10K to 60K, 
the \emph{FSC} decreases, while the \emph{ARE} increases slowly.
It can be seen that using pipelined tables with a weight around $\alpha=0.7$ achieves the best result.
Compared with a multi-hash table, pipelined tables can improve the \emph{FSC} by 3.1\%, 
and reduce the \emph{ARE} by 37.3\% respectively. 
This confirms our theoretical analysis on $\alpha$ in Section \ref{analysis}.
In the experiments thereafter, we will use a default weight of 0.7.

In Fig. \ref{fig:comparison_increase_depth}, we plot the \emph{Average Relative Error (ARE)} 
for flow size estimation of 50K flows, when the depth of the main table is set to 1, 2, 3 and 4.
It can be seen that increasing $d$ from 1 to 3 reduces the \emph{ARE} by around 3 times (i.e., from 0.34 to 0.12), 
while increasing $d$ from 3 to 4 will have only a minor improvement (i.e., from 0.12 to 0.075). 
In the experiments thereafter, we will use a default depth of 3.


\subsection{Application Performance }
In this section, we evaluate the performance of HashFlow against HashPipe, ElasticSketch and FlowRadar, 
for typical measurement applications as described in  Section~\ref{methodology}.

\begin{figure*}[ht!]
    \centering
    \mbox{
    \subfigure[CAIDA\label{subfig:caidaflowmonitoring}]{\includegraphics[width=0.24\linewidth]{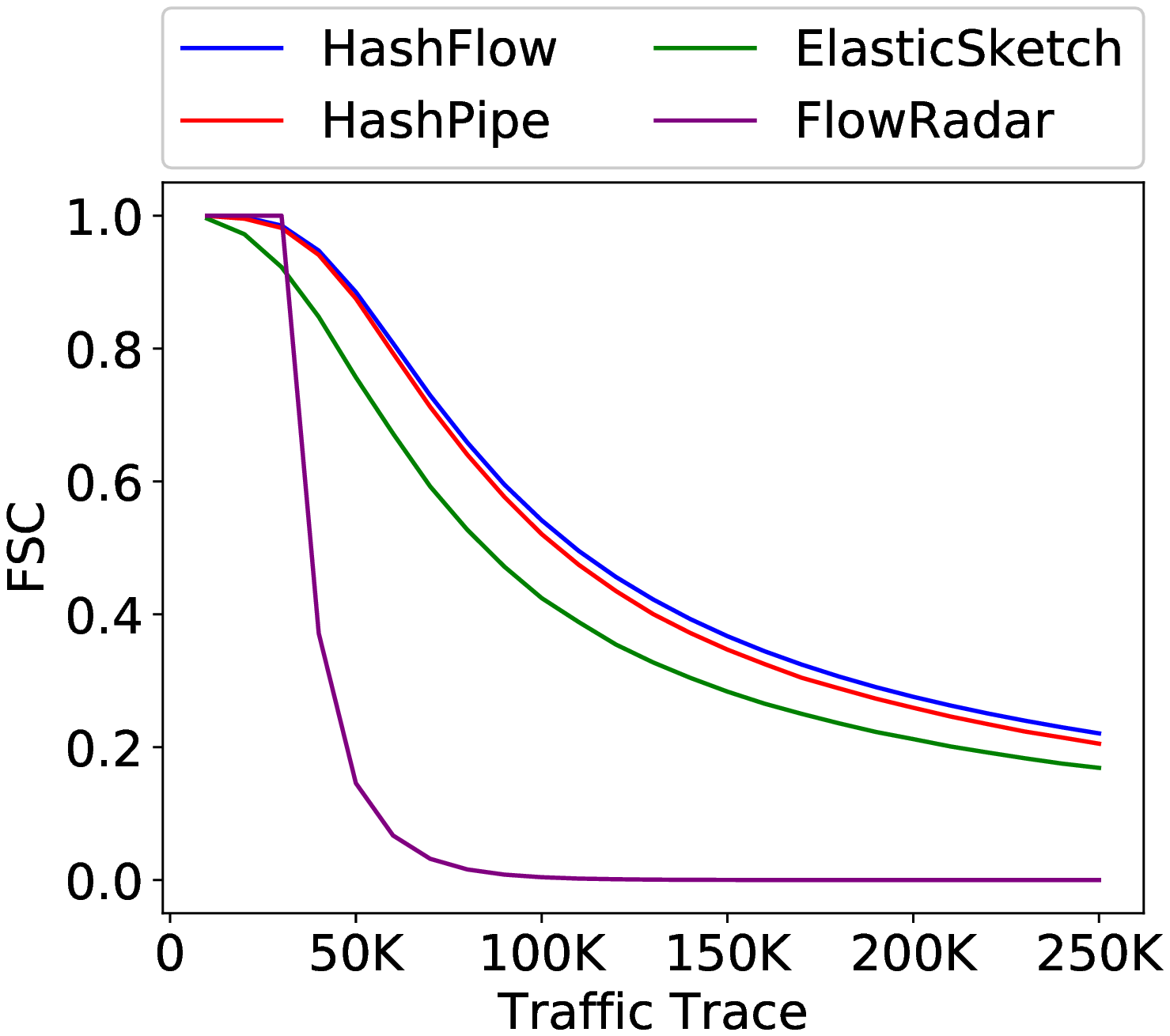}}
    \subfigure[Campus Network\label{subfig:campusnetworkflowmonitoring}]{\includegraphics[width=0.24\linewidth]{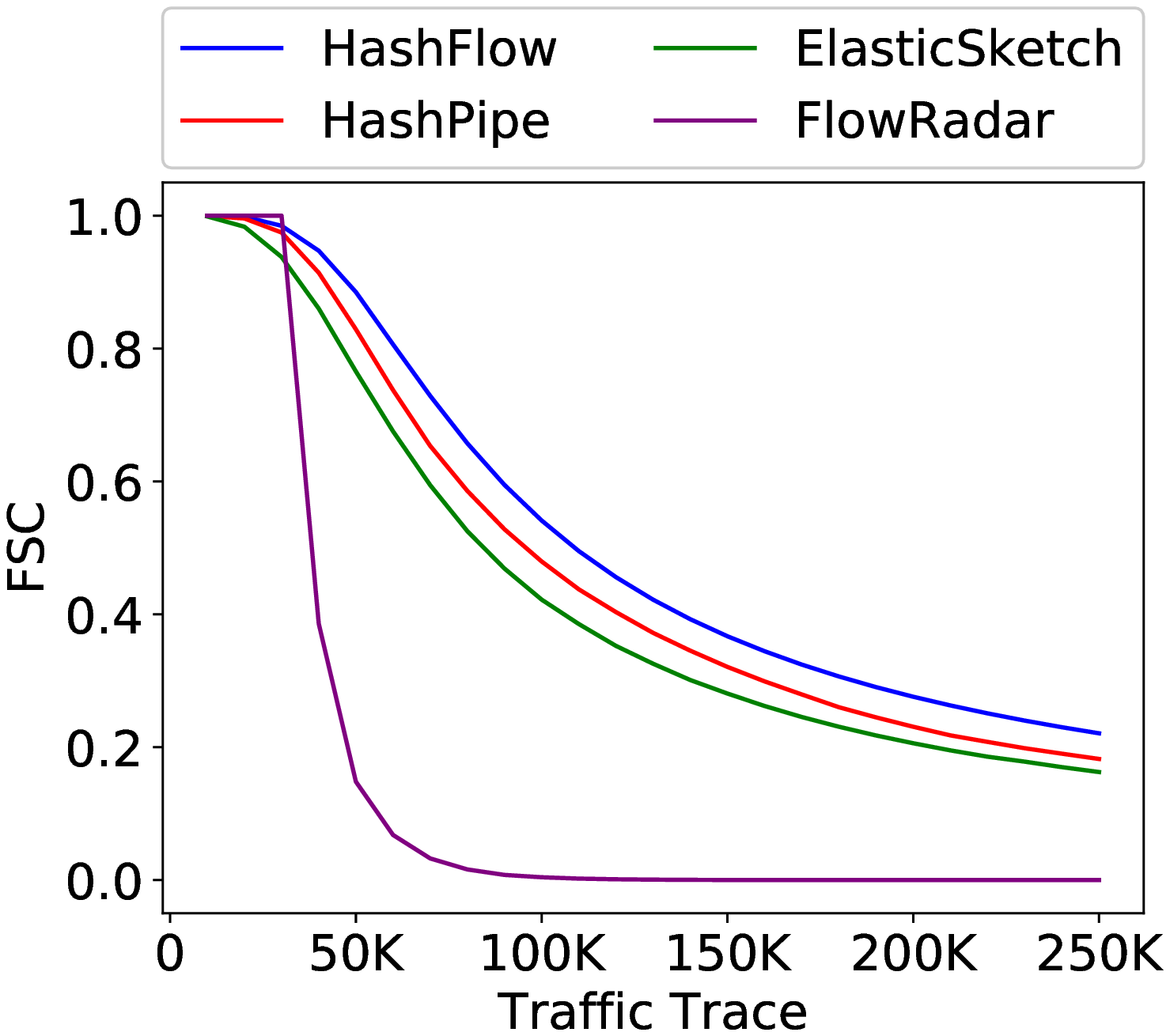}}
    \subfigure[ISP1\label{subfig:hgcflowmonitoring}]{\includegraphics[width=0.24\linewidth]{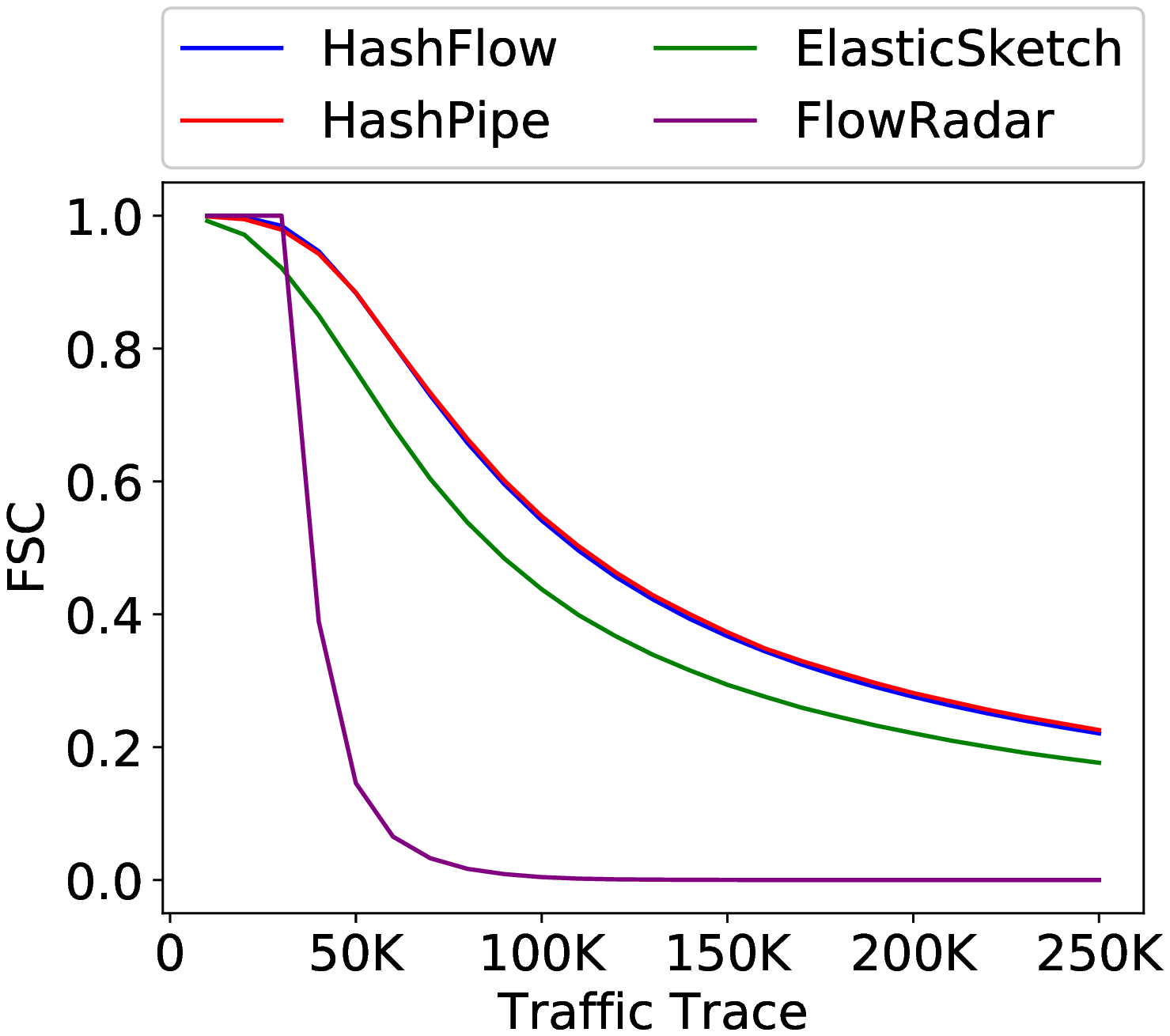}}
    \subfigure[ISP2\label{subfig:telecomflowmonitoring}]{\includegraphics[width=0.24\linewidth]{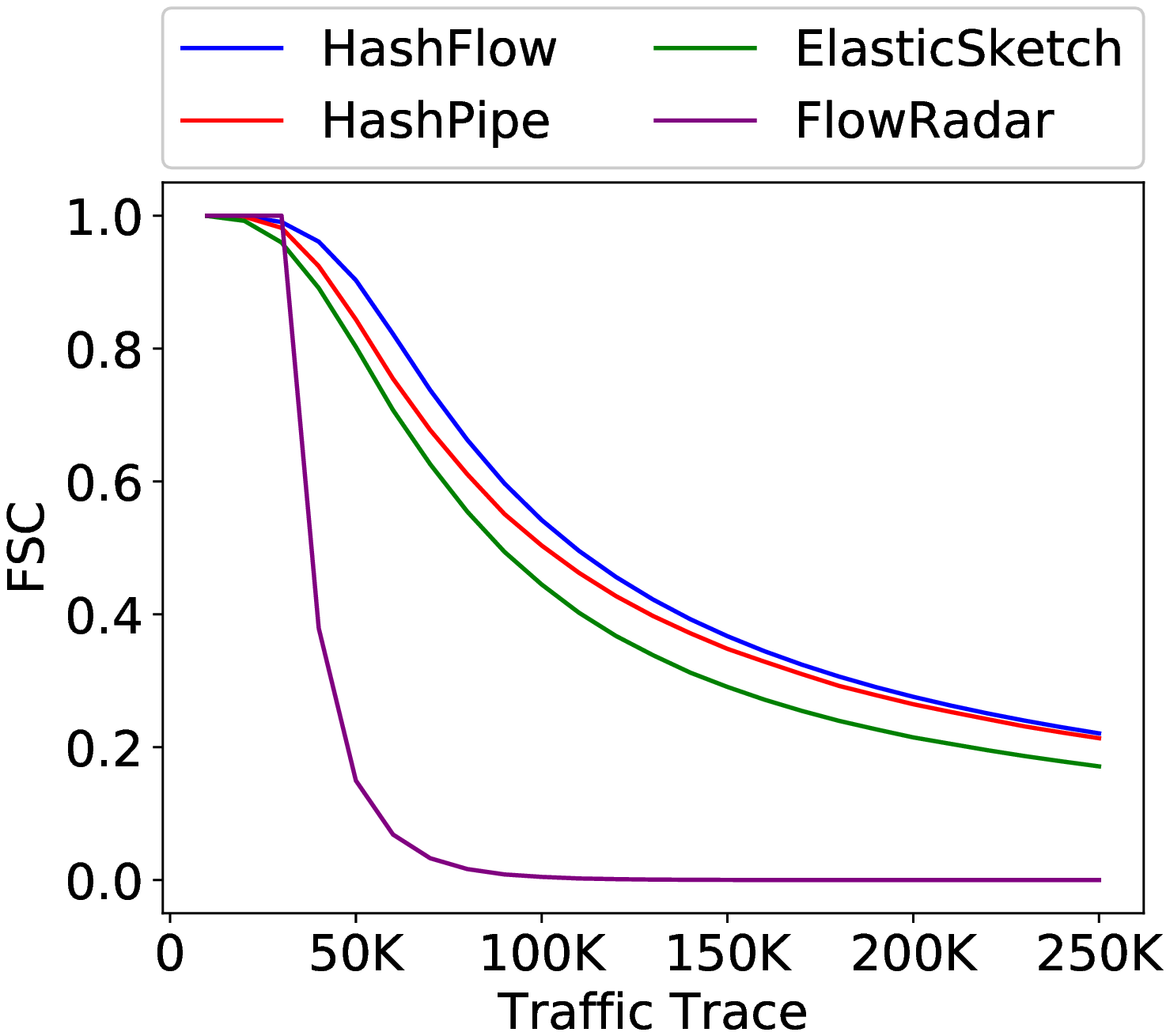}}
    }
    \caption{\emph{Flow Set Coverage (FSC)} for \emph{Flow Record Report}}
    \label{fig:comparison_concurrent_flows_increases_flow_monitoring}
\end{figure*}

Fig.~\ref{fig:comparison_concurrent_flows_increases_flow_monitoring} depicts the 
\emph{Flow Set Coverage (FSC)} for flow record report achieved by these algorithms.
We can see HashFlow nearly always performs better than the others. For example, 
for a total of 250K flows, it can successfully report around 55K flows, 
nearly making a full use of its main table. 
Its \emph{FSC} is more than 20\% higher than ElasticSketch in all traces, 
and is that higher than HashPipe in the Campus Network trace. 
The only exception when HashFlow loses is that, 
for a very small number of flows (the left up corner in the figures), 
FlowRadar has the highest coverage. 
This is because FlowRadar can successfully decode nearly all flow records when only a few flows arrive. 
But its performance drops significantly soon after the flow count goes beyond a certain point, 
since after that, too many flows mixed up, and the decoding often fails.

\begin{figure*}[ht!]
    \centering
    \mbox{
        \subfigure[CAIDA\label{subfig:caidacardinality}]{\includegraphics[width=0.24\linewidth]{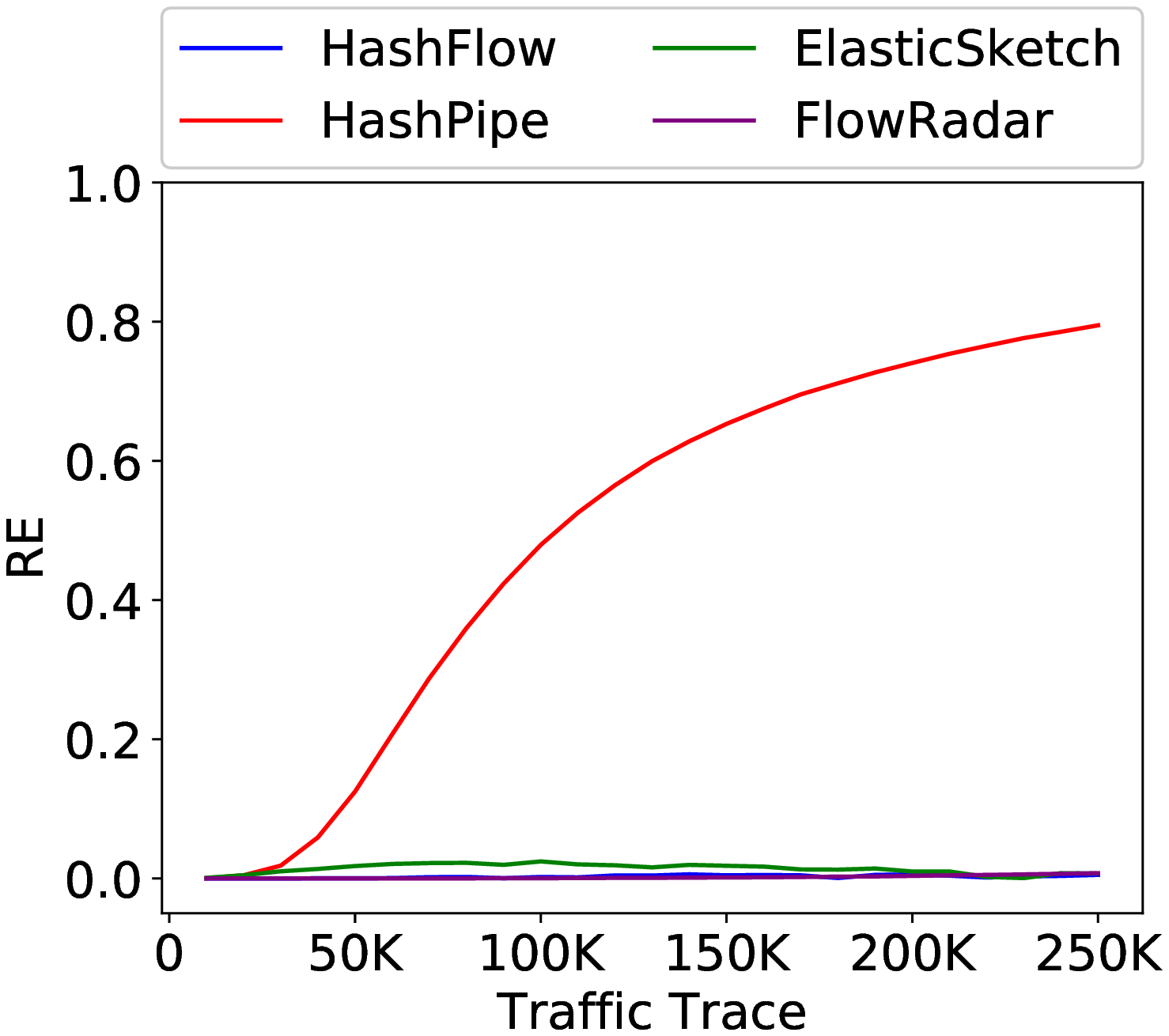}}
        \subfigure[Campus Network\label{subfig:campusnetworkcardinality}]{\includegraphics[width=0.24\linewidth]{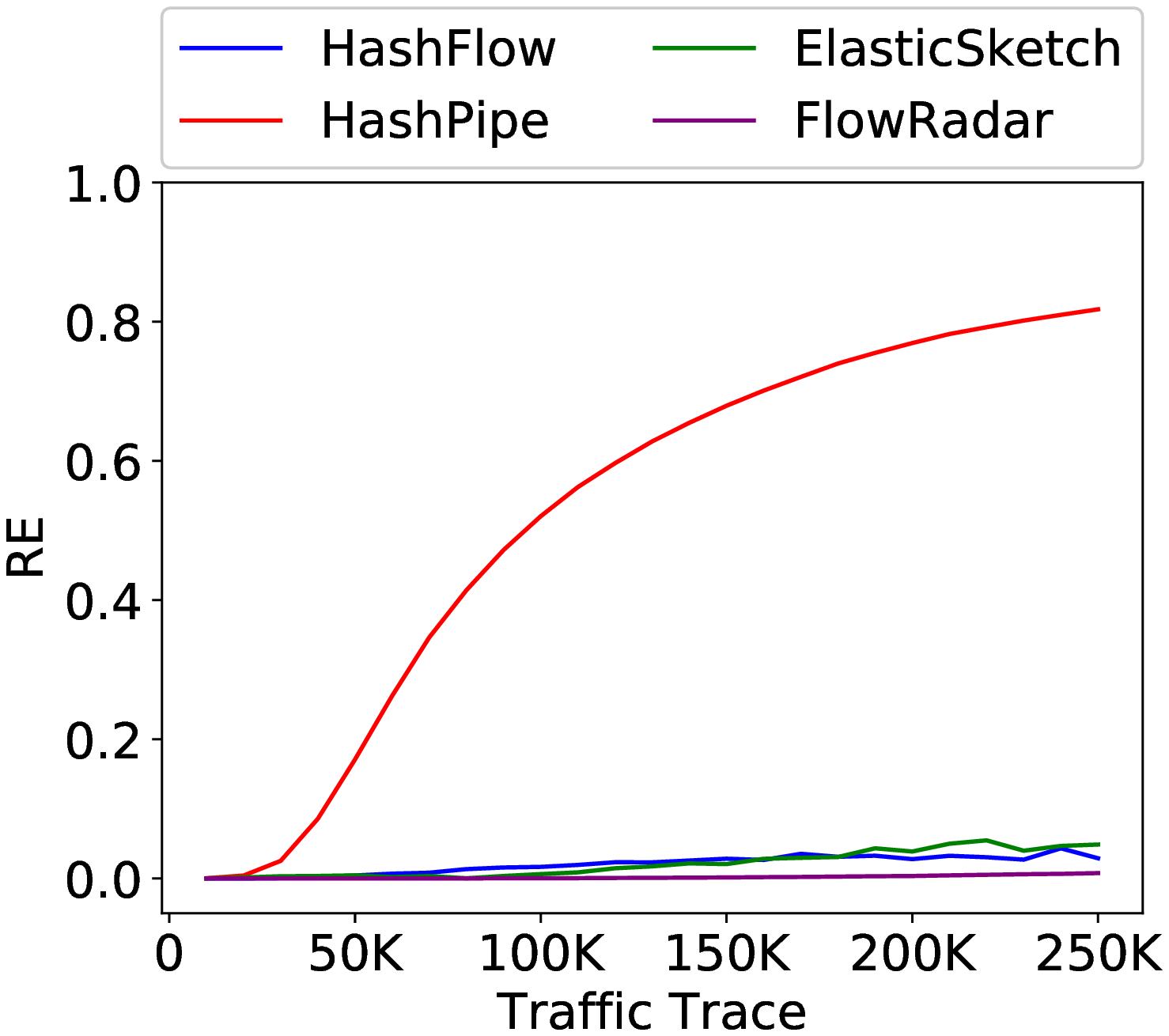}}
        \subfigure[ISP1\label{subfig:hgccardinality}]{\includegraphics[width=0.24\linewidth]{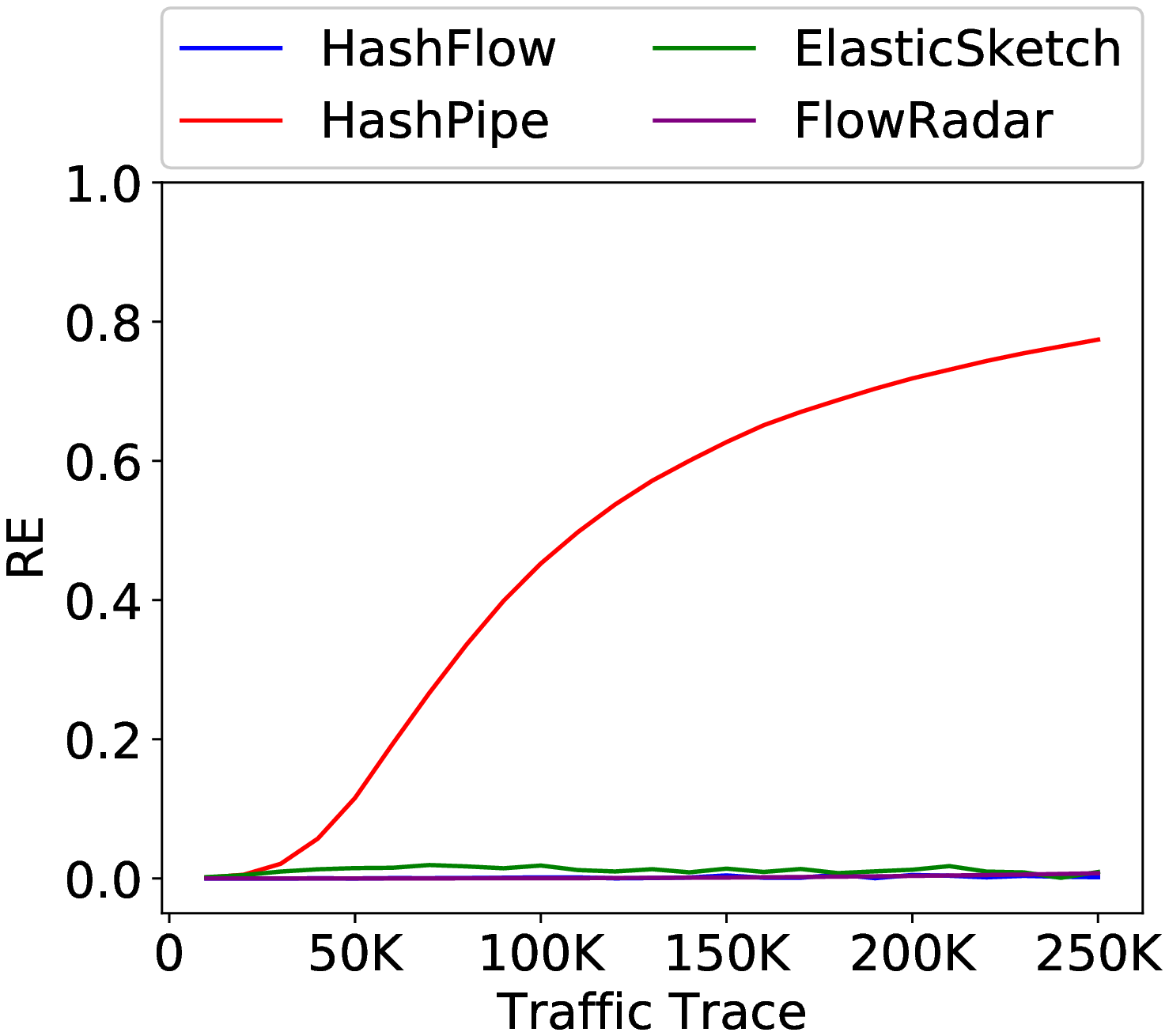}}
        \subfigure[ISP2 Trace\label{subfig:telecomcardinality}]{\includegraphics[width=0.24\linewidth]{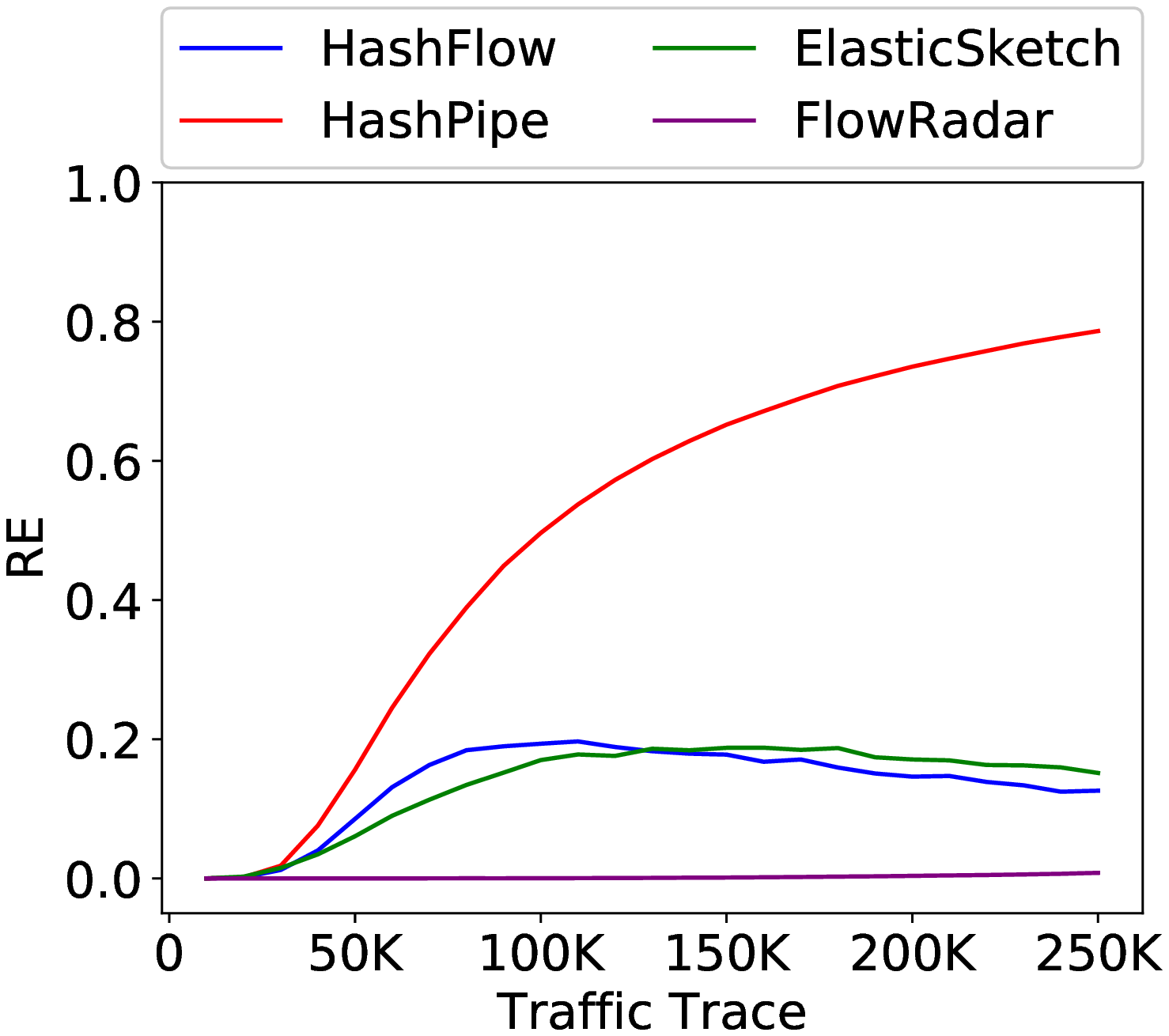}}
    }
    \caption{\emph{Relative Error (RE)} for \emph{Flow Cardinality Estimation}}
    \label{fig:comparison_concurrent_flows_increases_cardinality}
\end{figure*}

Fig.~\ref{fig:comparison_concurrent_flows_increases_cardinality} shows the results 
of estimating the total number of flows, where in most of the time, 
HashFlow, ElasticSketch and FlowRadar achieve a similar level of accuracy. 
Among them, FlowRadar works slightly better since it uses a bloom filter to count flows, 
which is not sensitive to flow sizes, while HashFlow and ElasticSketch are slightly affected 
by the flow size distribution due to their assumption on the existence of elephant and mice flows. 
This is particularly true in the ISP2 trace, where nearly all flows contain less than 5 packets.
HashPipe always performs badly since it does not use any advanced cardinality estimation technique 
to compensate for the flows it drops. 

\begin{figure*}[ht!]
    \centering
    \mbox{
        \subfigure[CAIDA Trace\label{subfig:caidafsestimation}]{\includegraphics[width=0.24\linewidth]{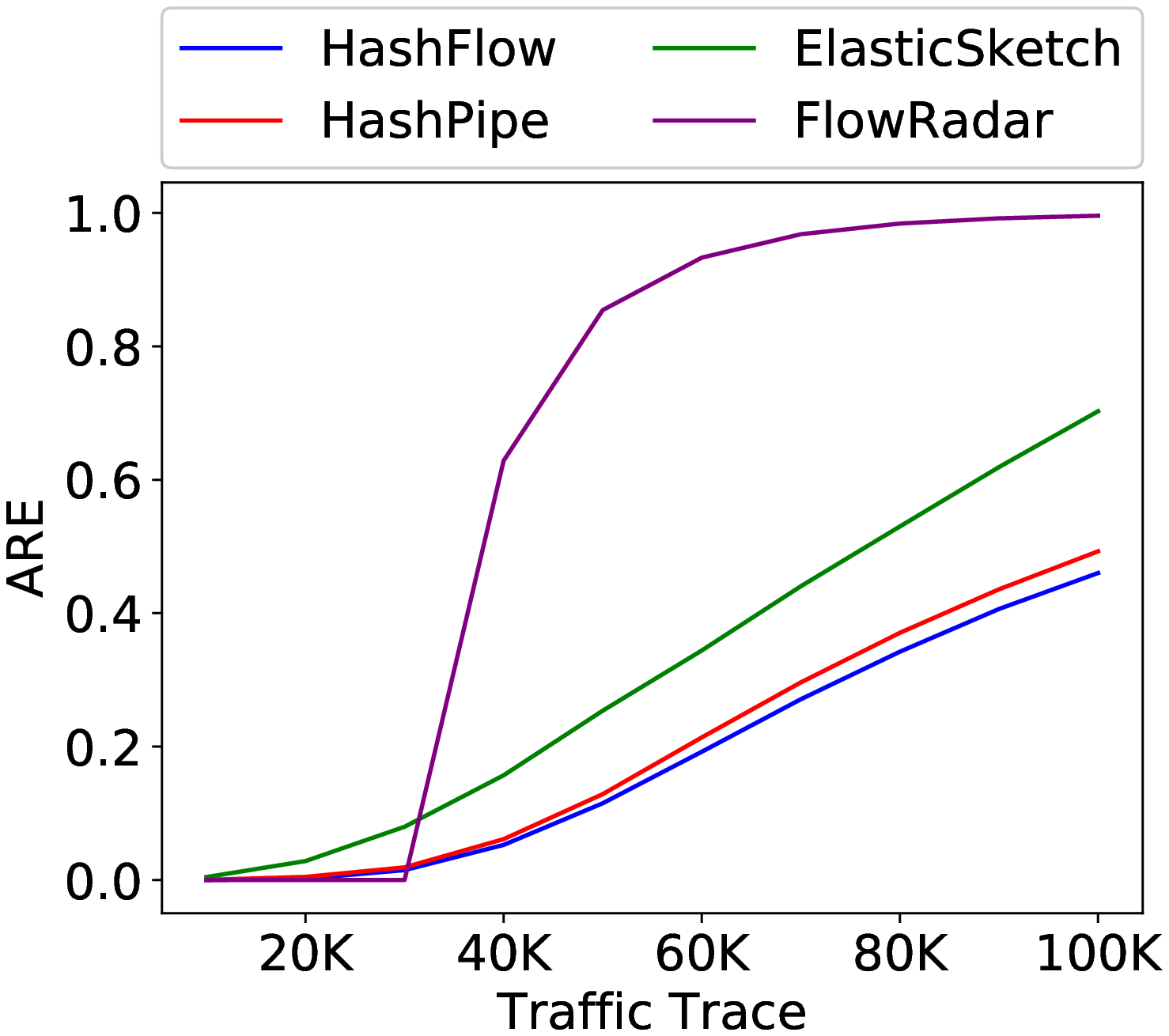}}
        \subfigure[Campus Network Trace\label{subfig:campusnetworkfsestimation}]{\includegraphics[width=0.24\linewidth]{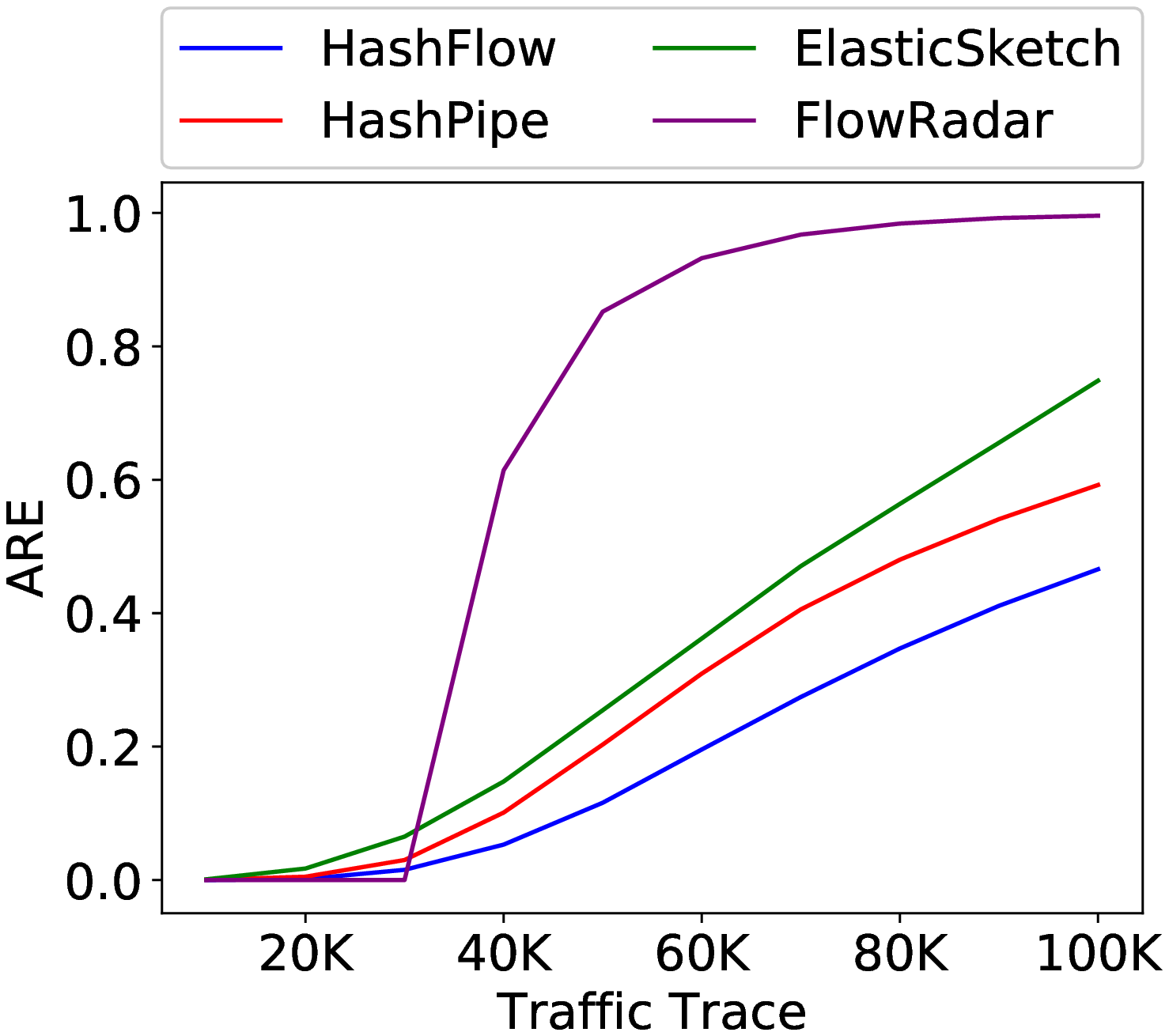}}
        \subfigure[HGC Trace\label{subfig:hgcfsestimation}]{\includegraphics[width=0.24\linewidth]{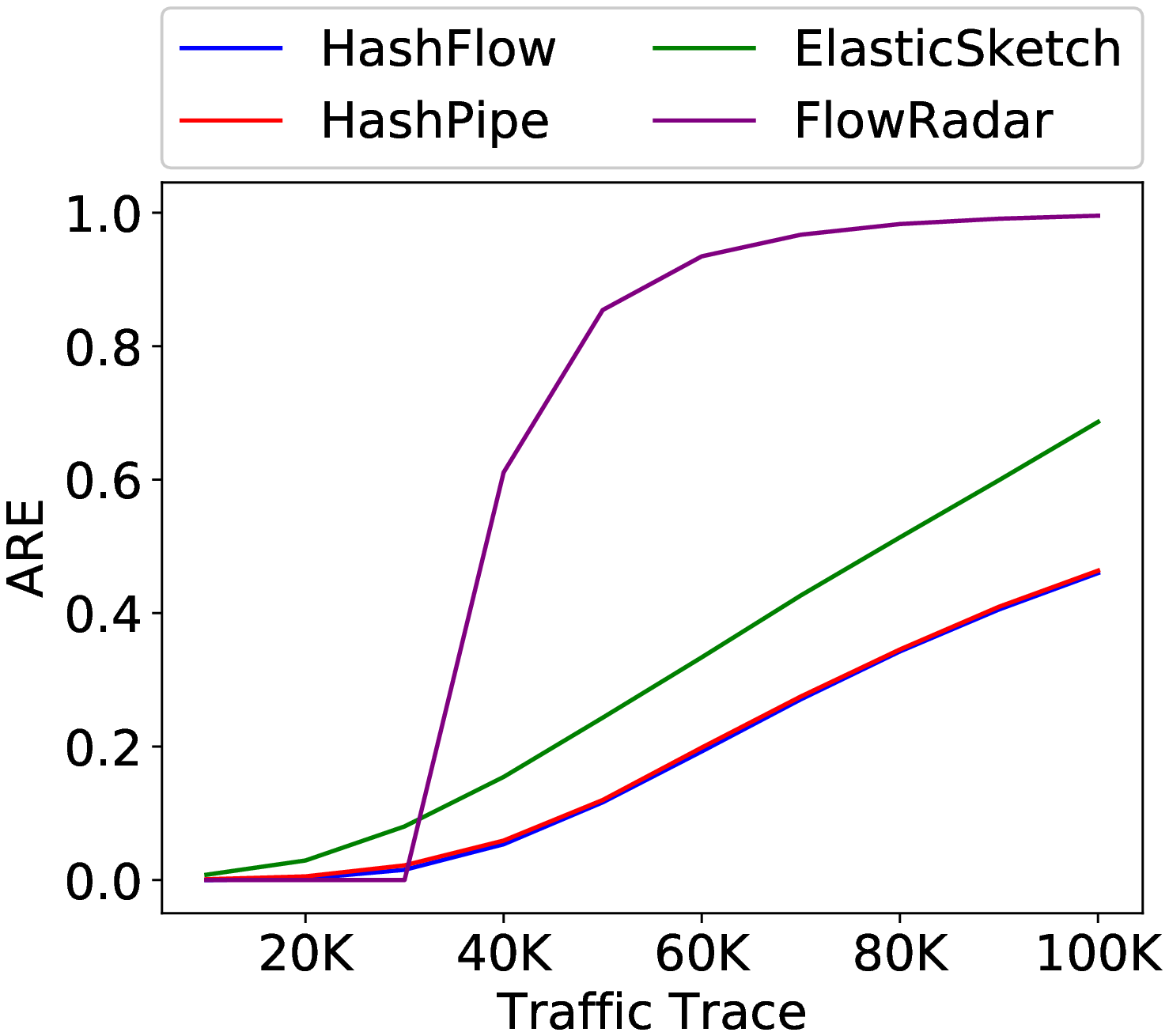}}
        \subfigure[Telecom Trace\label{subfig:telecomfsestimation}]{\includegraphics[width=0.24\linewidth]{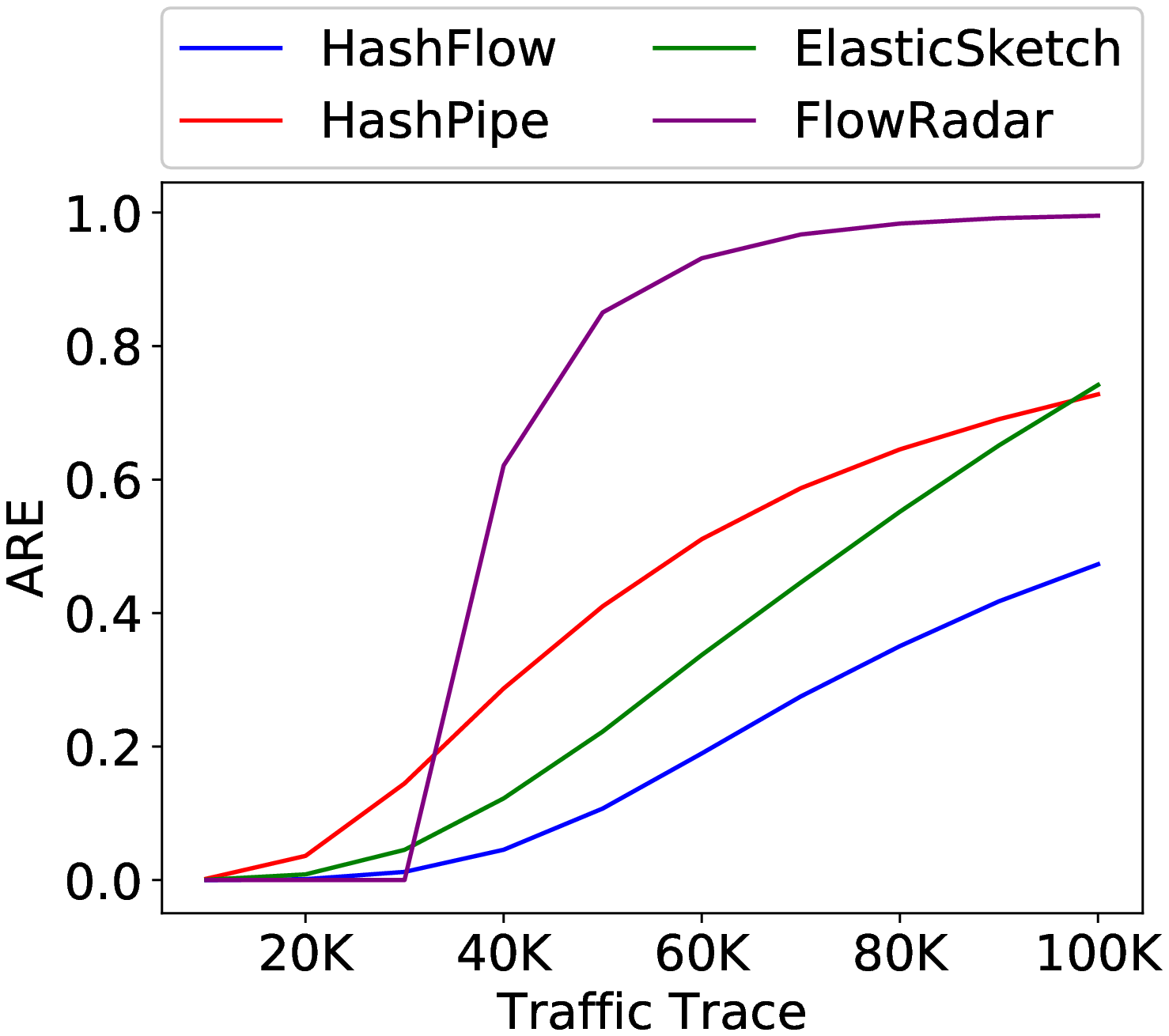}}
    }
    \caption{\emph{Average Relative Error (ARE)} for \emph{Flow Size Estimation}}
    \label{fig:comparison_concurrent_flows_increases_fs_estimation}
\end{figure*}

Fig. \ref{fig:comparison_concurrent_flows_increases_fs_estimation} shows how accurate these algorithms
can estimate the flow sizes.  HashFlow often achieves a much lower estimation error 
than its competitors. For example, when there are 100K flows, the relative estimation error
of HashFlow is around 0.4, while the error of the others is more than 0.6 (50\% higher) in most cases. 
FlowRadar performs very badly when there are more than 40K flows, 
while the accuracy of HashPipe is not very stable.

\begin{figure*}[ht!]
    \centering
    \mbox{
        \subfigure[CAIDA\label{subfig:caidahhdf1score}]{\includegraphics[width=0.24\linewidth]{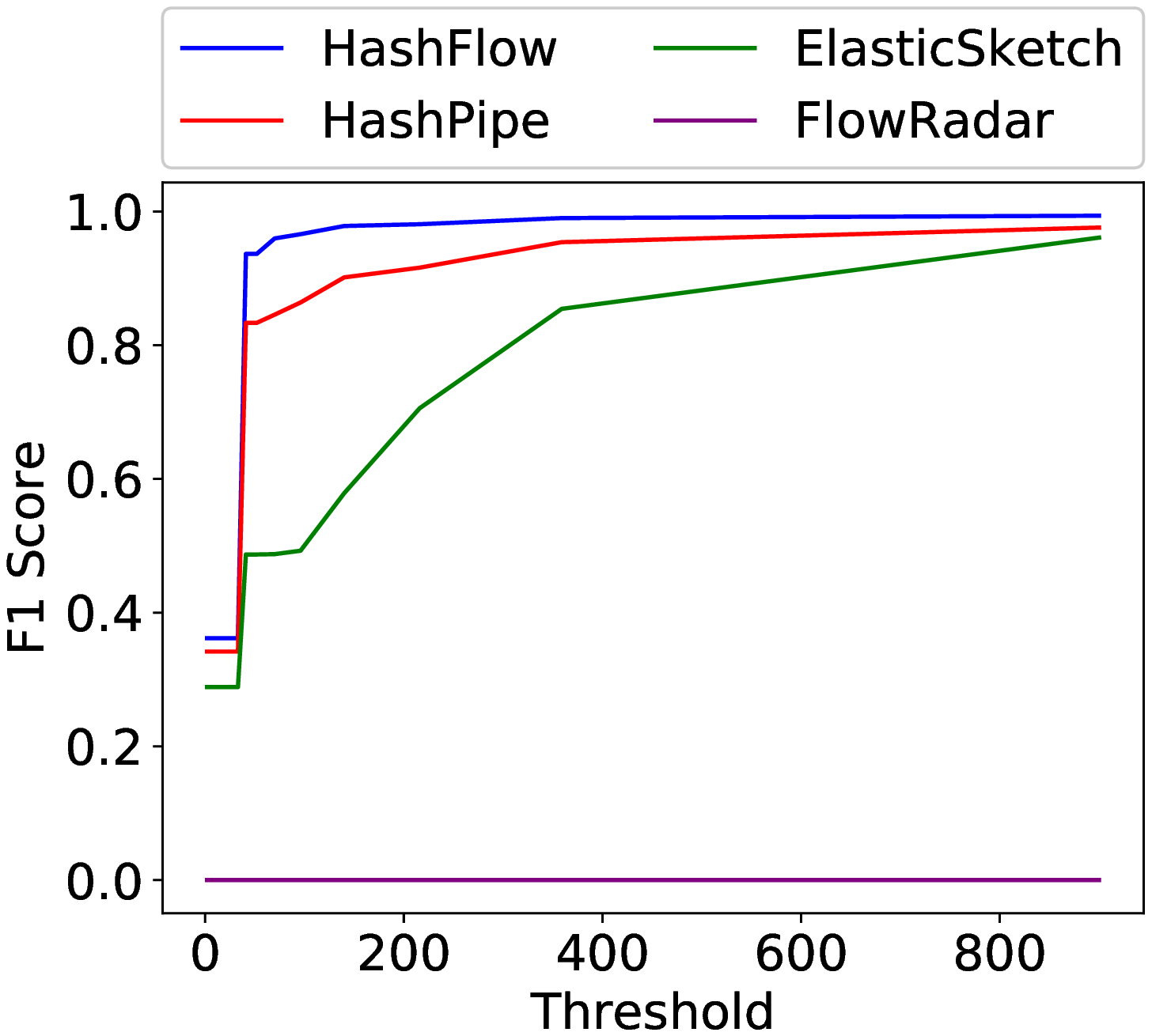}}
        \subfigure[Campus Network\label{subfig:campusnetworkhhdf1score}]{\includegraphics[width=0.24\linewidth]{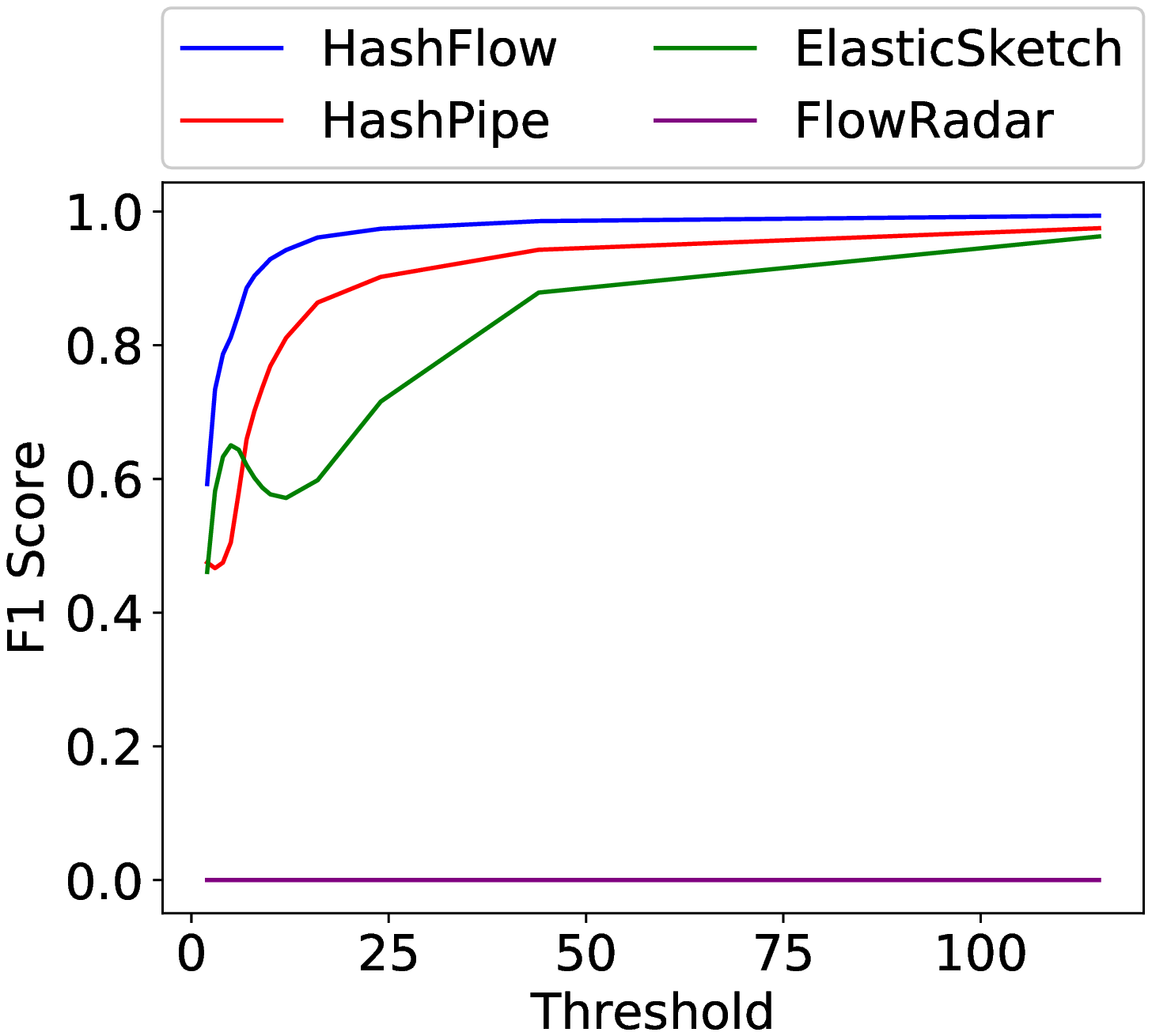}}
        \subfigure[ISP1\label{subfig:hgchhdf1score}]{\includegraphics[width=0.24\linewidth]{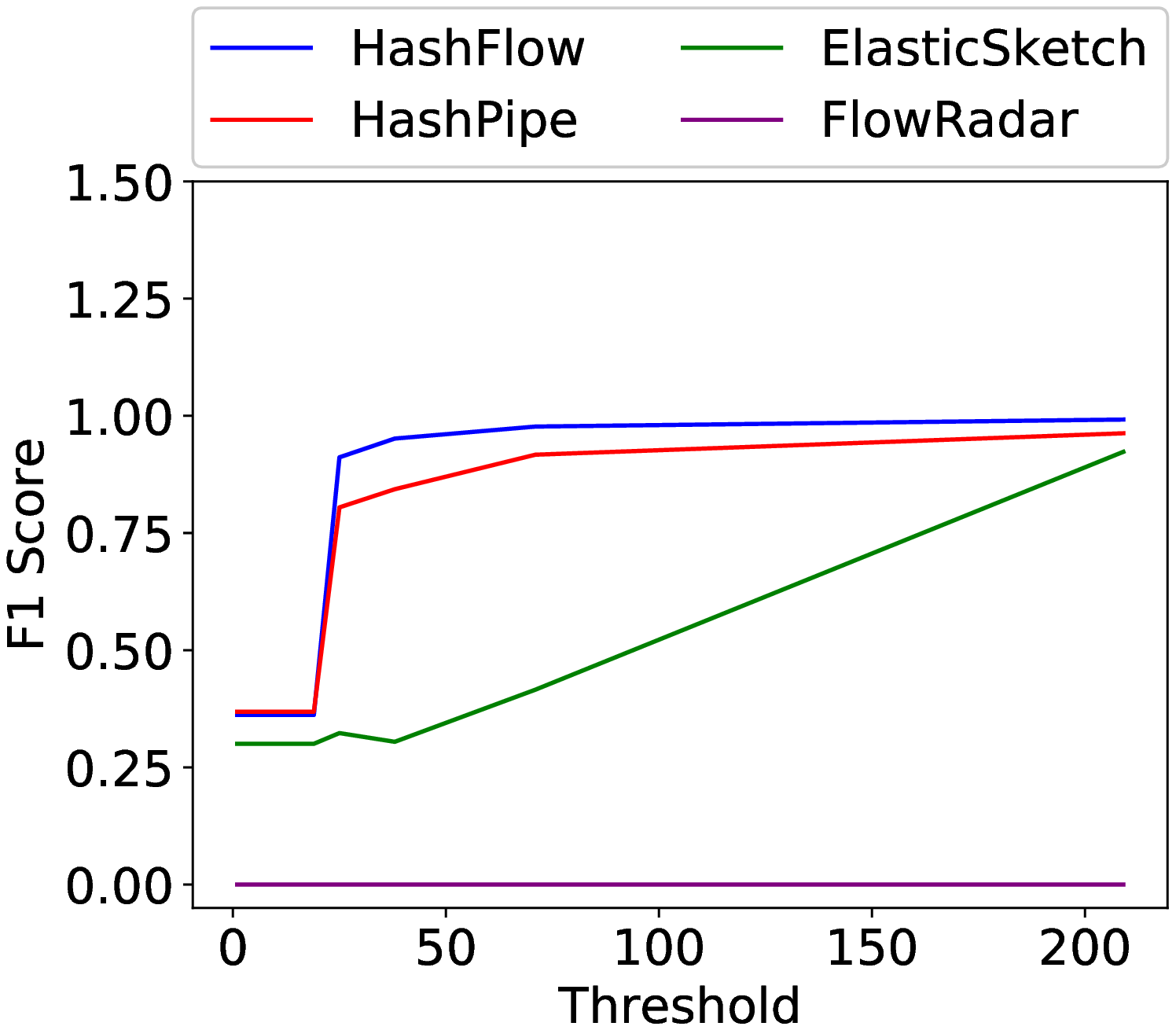}}
        \subfigure[ISP2\label{subfig:telecomhhdf1score}]{\includegraphics[width=0.24\linewidth]{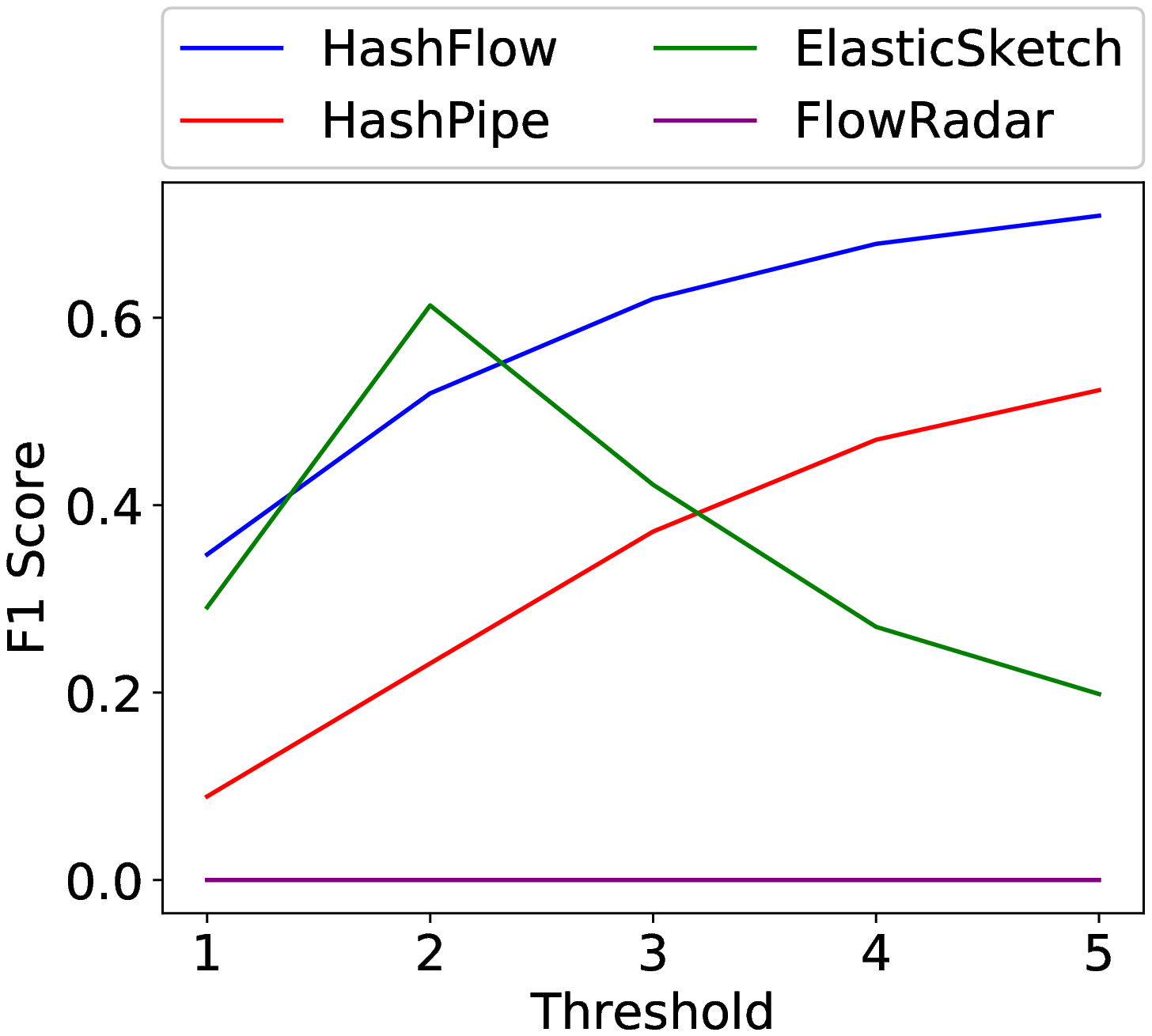}}
    }
    \caption{\emph{F1 Score} for \emph{Heavy Hitter Detection}}
    \label{fig:comparison_concurrent_flows_increases_hhd_f1_score}
\end{figure*}

\begin{figure*}[ht!]
    \centering
    \mbox{
        \subfigure[CAIDA\label{subfig:caidahhdare}]{\includegraphics[width=0.24\linewidth]{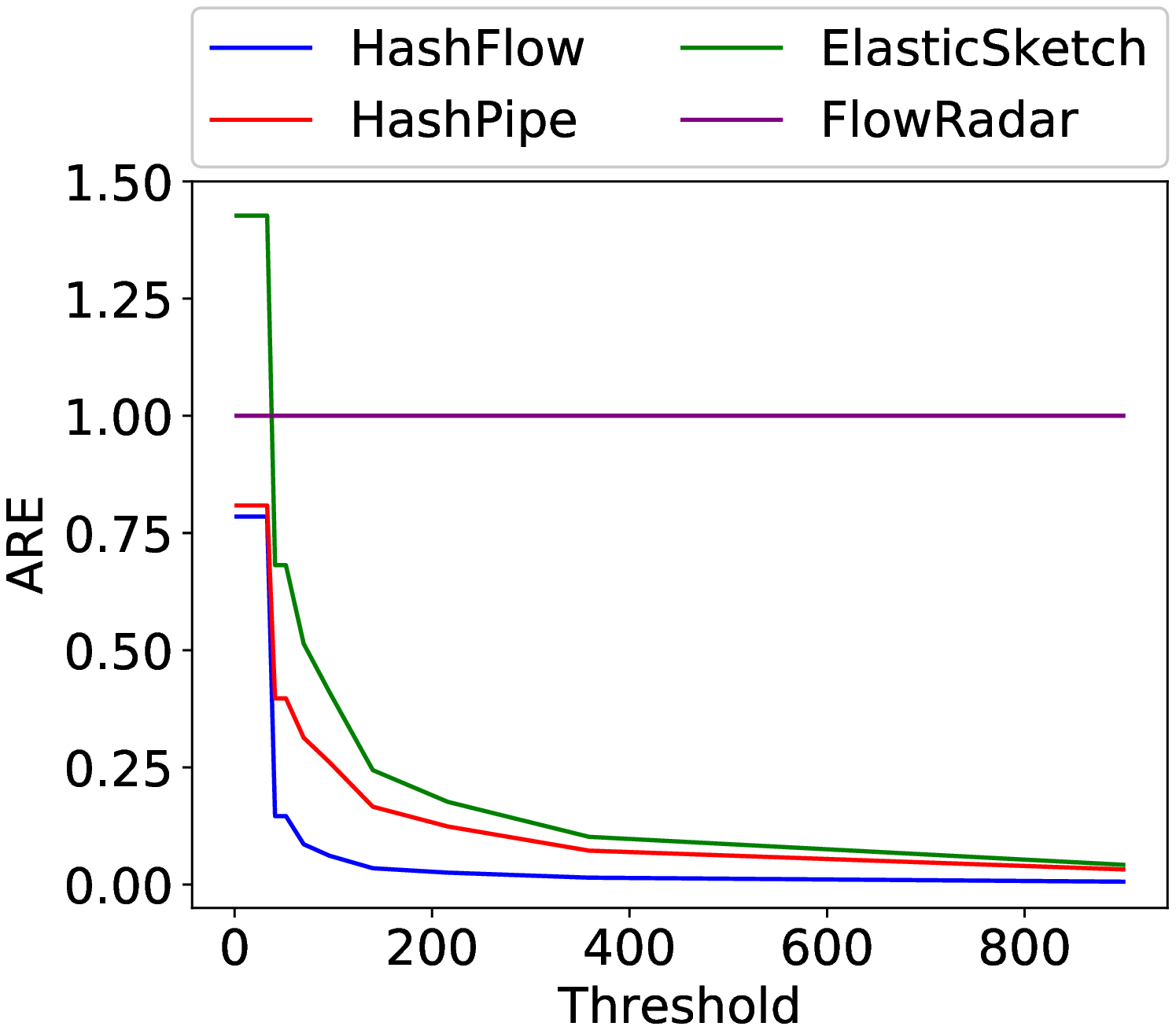}}
        \subfigure[Campus Network\label{subfig:campusnetworkhhdare}]{\includegraphics[width=0.24\linewidth]{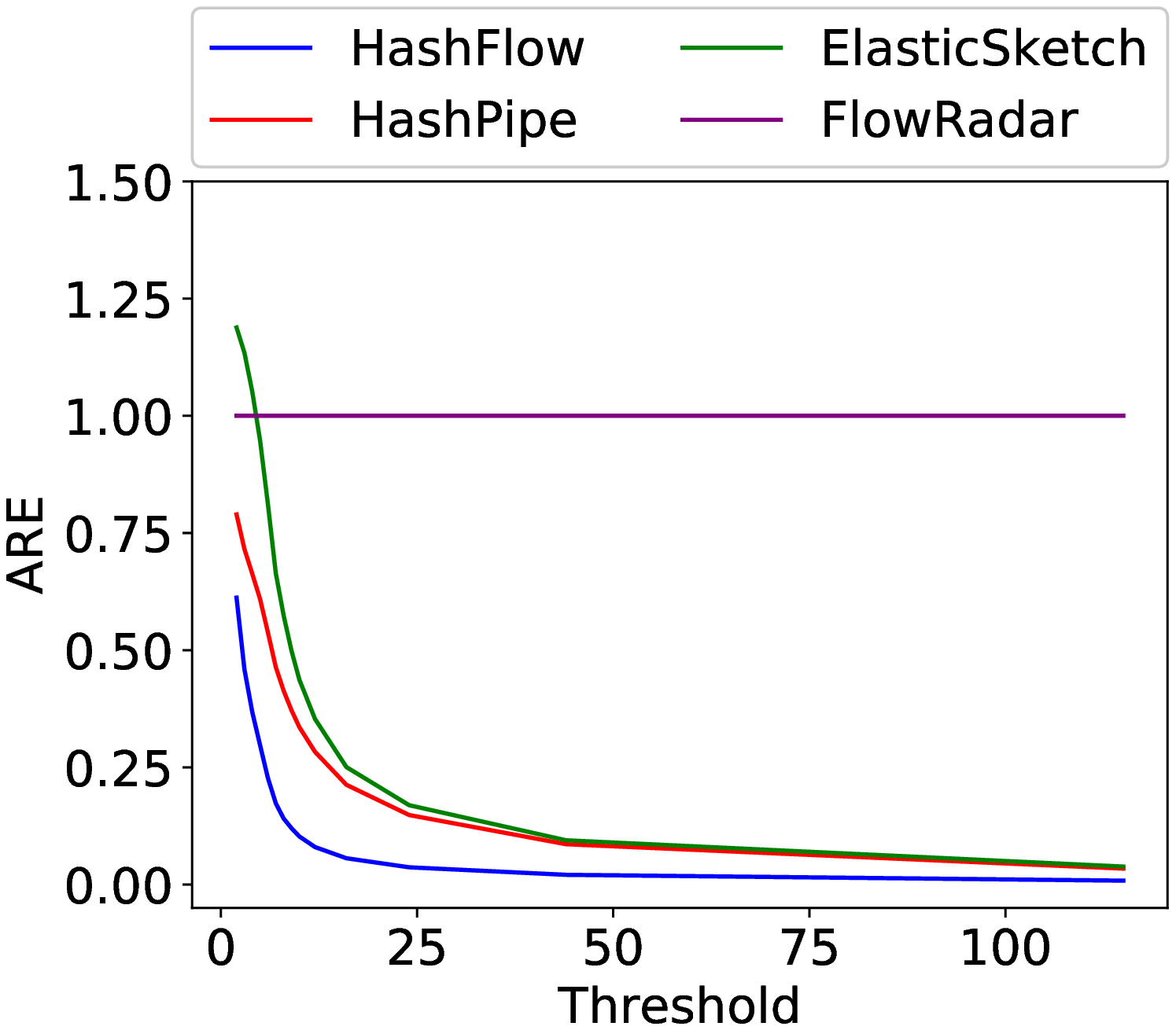}}
        \subfigure[ISP1\label{subfig:hgchhdare}]{\includegraphics[width=0.24\linewidth]{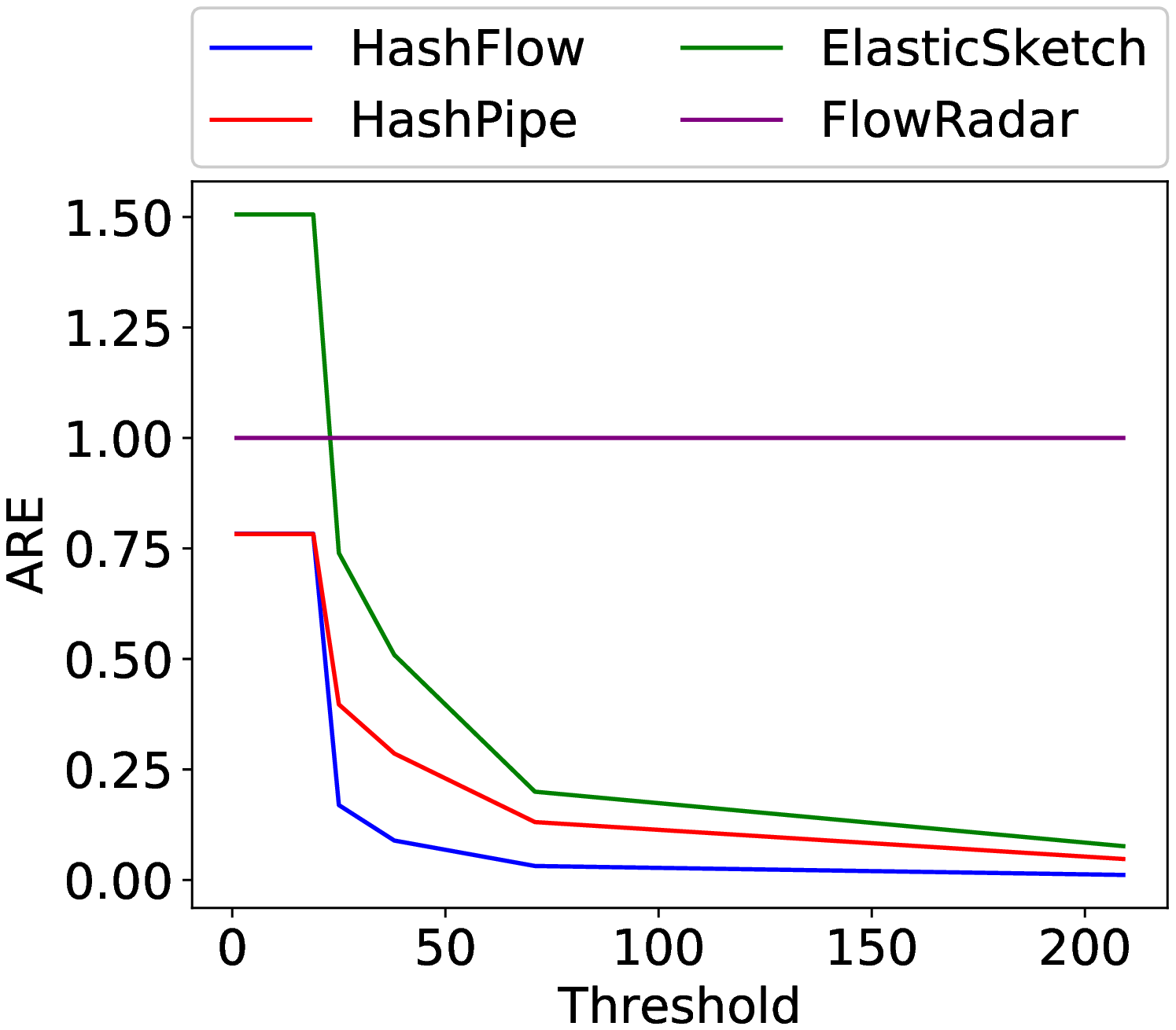}}
        \subfigure[ISP2\label{subfig:telecomhhdare}]{\includegraphics[width=0.24\linewidth]{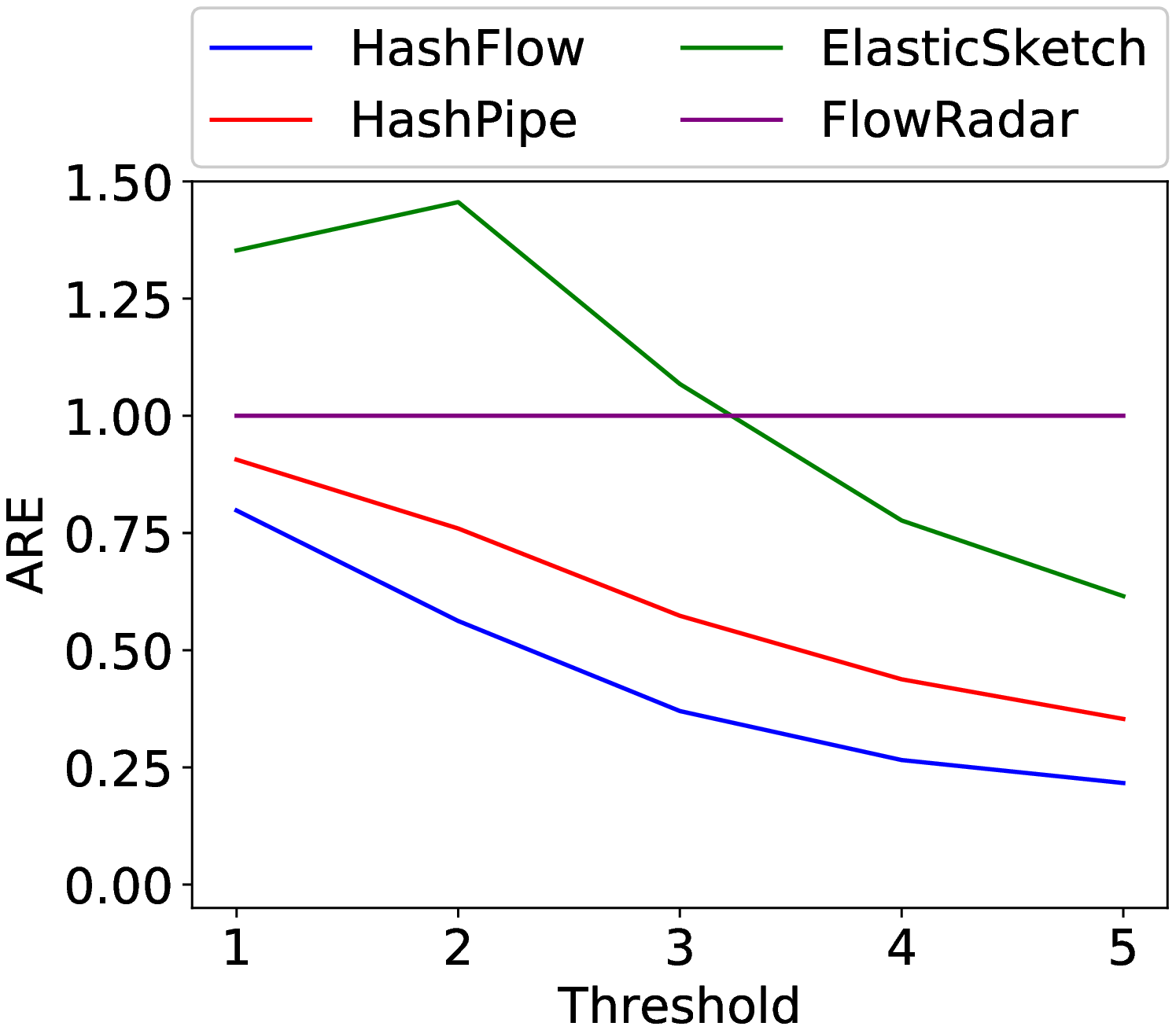}}
    }
    \caption{\emph{Average Relative Error(ARE)} for \emph{Heavy Hitter Detection}}
    \label{fig:comparison_concurrent_flows_increases_hhd_are}
\end{figure*}

At last, we show whether they can accurately detect heavy hitters.
We feed 250K flows to each algorithm, and measure their capabilities by the \emph{F1 Score} of their detection accuracy, 
where the threshold for a flow to be treated as a heavy hitter varies. 
We also measure their $ARE$ when estimating the sizes of the detected heavy hitters.
The results are depicted in Fig. \ref{fig:comparison_concurrent_flows_increases_hhd_f1_score} 
and Fig.~\ref{fig:comparison_concurrent_flows_increases_hhd_are}, respectively. 
Apparently, FlowRadar is not a good candidate under such heavy load. 
HashPipe is designed specifically for detecting heavy hitters, 
but our HashFlow still outperforms it in nearly all cases, for both metrics. 
Not considering the extreme case of the ISP2 trace where most flows are typically very small, 
for a wide range of thresholds, HashFlow achieves a \emph{F1 Score} of 1 (accurately detecting all heavy hitters) 
when the scores of HashPipe and ElasticSketch are around 0.9 and 0.4 $\sim$ 0.7, respectively. 
On the other hand, when HashFlow makes nearly perfect size estimation of the heavy hitters, 
the \emph{ARE} of HashPipe and ElasticSketch are around 0.15 $\sim$ 0.2 and 0.2 $\sim$ 0.25, respectively.
Even with a very small threshold used in the ISP2 trace, HashFlow still clearly outperforms the others.

\subsection{Throughput}
\begin{figure*}[ht!]
    \centering
    \mbox{
        \subfigure[Throughput\label{fig:throughput}]{\includegraphics[width=0.25\linewidth]{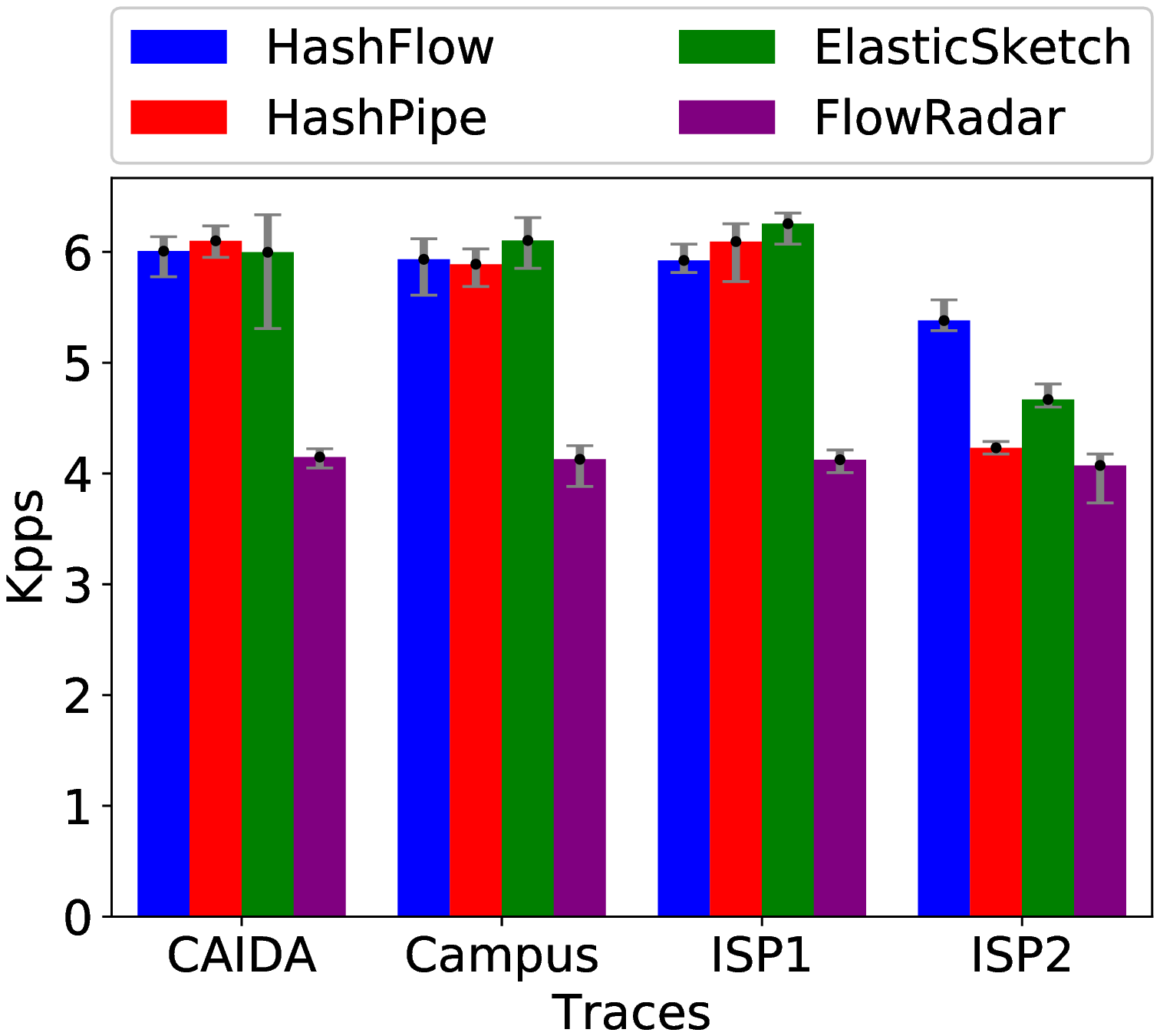}}
        \subfigure[Number of hash operations\label{fig:avehash}]{\includegraphics[width=0.25\linewidth]{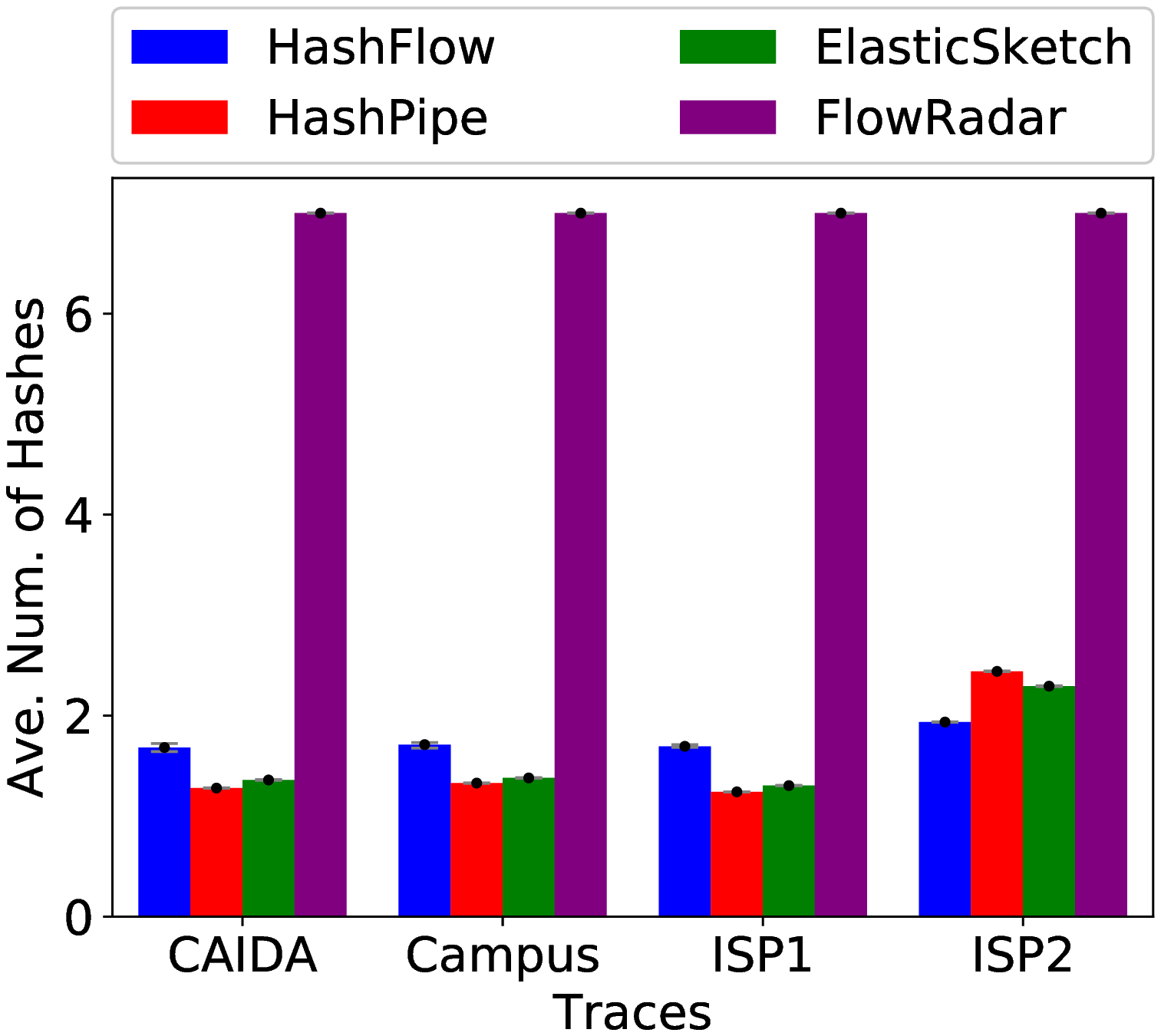}}
        \subfigure[Number of memory access\label{fig:avemem}]{\includegraphics[width=0.25\linewidth]{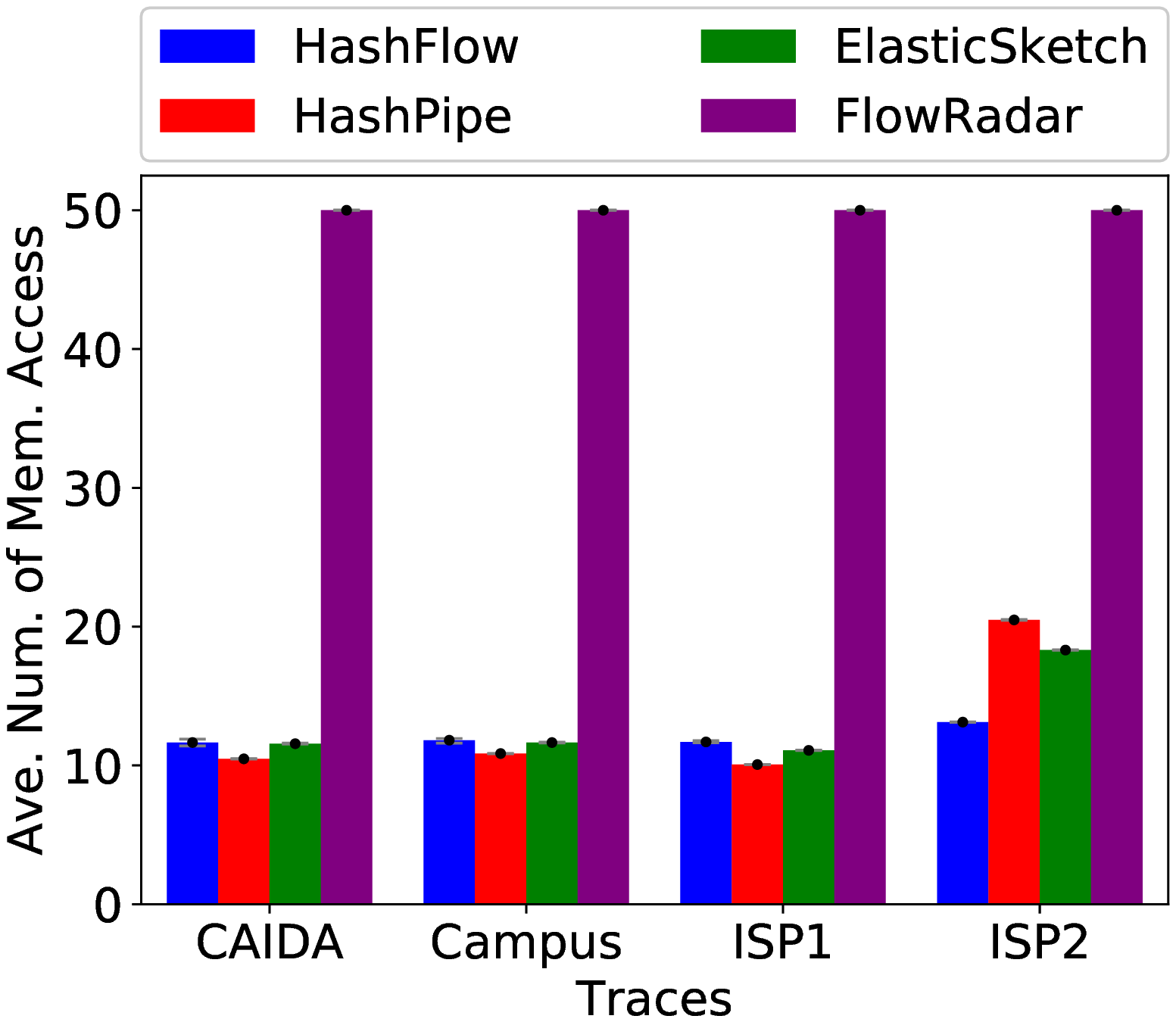}}
    }
    \caption{Throughput, Hash Operation and Memory Access}
    \label{throughput}
\end{figure*}

Since we don't have a hardware switch with P4 programmability, 
we test the throughput of these algorithms with bmv2, 
on a PC with Intel(R) Core(TM) i5-4680K CPU@3.40GHz, 
where each CPU core owns a 6144 KB cache. 
We  use \emph{isolcpus} to isolate the cores to prevent context switches.
Bmv2 achieves around 20 Kpps forwarding speed, 
and the throughput after loading the algorithms are depicted in Fig. \ref{fig:throughput}.
To obtain a better understanding, we also record the average number of 
hash operations, as well as memory accesses, for each algorithm. 
The  results in Fig. \ref{fig:avehash} and Fig. \ref{fig:avemem} indicate that, 
even if the throughput on a software switch is not convincing, 
HashFlow will perform comparably to HashPipe and ElasticSketch, 
and much better than FlowRadar.

\section{Conclusion}
\label{section:conclusion}
We propose HashFlow for efficient collection of flow records, 
which are useful for a wide range of measurement and analysis applications. 
The collision resolution and record promotion strategy is of central importance 
to HashFlow's accuracy and efficiency. We analyze the performance bound 
of HashFlow based on a probabilistic model, and implement it in a software switch. 
The evaluation results based on real traces from different networks show that, 
HashFlow consistently achieves a clear better performance in nearly all cases. 
This is due to its high utilization of memory with only few operations.
In the future, we plan to port it to real hardware switches, 
and study how to make it adaptive to traffic variation and network wide measurement.

\bibliographystyle{IEEEtran}
\bibliography{ms}
\end{document}